\documentclass[twocolumn,showpacs,amsfonts]{revtex4-1}

\usepackage{amsmath}
\usepackage{latexsym}
\usepackage{float}
 \usepackage{amssymb}
\usepackage{graphicx}
\usepackage{textcomp}
\usepackage{hyperref}
\usepackage{subfigure}
\usepackage{color}
\def\ba{\begin{eqnarray}}
\def\ea{\end{eqnarray}}
\def\be{\begin{equation}}
\def\ee{\end{equation}}
\def\bm{\begin{math}}
\def\me{\end{math}}

\newcommand{\dummy}

\begin{document}
\title{How Do Clusters in Phase-Separating Active Matter Systems Grow?
A study for Vicsek activity in systems undergoing  vapor-solid transition}
\author{ Subhajit Paul$^{1,2}$, Arabinda Bera$^{1,3}$, and Subir K. Das$^{1,3,*}$}
\affiliation{$^1$ Theoretical Sciences Unit, Jawaharlal Nehru Centre for Advanced Scientific Research,
 Jakkur P.O, Bangalore 560064, India}
 \affiliation{$^2$ Instit\"ut f\"ur Theoretische Physik, Universit\"at Leipzig, IPF 231101, 04081, Leipzig, Germany}
\affiliation{$^3$ School of Advanced Materials, Jawaharlal Nehru Centre for Advanced Scientific Research,
 Jakkur P.O, Bangalore 560064, India}

\date{\today}

\begin{abstract}
~~Via molecular dynamics simulations we have studied kinetics of vapor-``solid'' phase transition in an
active matter model in which self-propulsion is introduced via the well-known Vicsek rule. The overall density 
of the particles is chosen
in such a way that the evolution morphology consists of disconnected clusters 
that are defined as regions of high density of particles. 
Our focus has been on understanding the influence of the above mentioned
self-propulsion on structure and growth of these clusters by comparing 
the results with those for the passive limit of the model
that also exhibits vapor-``solid'' transition. While in the passive case the growth occurs due to standard diffusive 
mechanism, the Vicsek activity leads to a very rapid growth, via a process
that is practically equivalent to the ballistic aggregation mechanism. 
The emerging growth law in the latter case has been accurately estimated and 
explained by invoking information on velocity and structural aspects of the clusters into a relevant theory. Some of these results are also discussed with reference to a model for the active Brownian particles.
\end{abstract}


\maketitle
\section{Introduction}
~There have been much research activities in the domain concerning kinetics in systems undergoing 
phase 
transitions \cite{onuki_6, binder1_6, bray_6, jones_6, lifs_6, amar_6, majum_6, binder2_6, binder3_6, siggia_6, 
furu1_6, furu2_6, roy_6, royp_6, shimi_6, mid1_6,marche_6, rama1_6, cates1_6, mishra1_6, belmon_6, mishra2_6, 
cates2_6, hagan1_6, hagan2_6, wys_6, mehes_6,
peru_6, mishra3_6, cremer_6, schwarz_6, pala_6, kumar1, das1_6, tref_6, vic1_6, czi_6, bagl_6, chate_6, das2_6, schak_6, capri_6, loi_6}. 
For conserved order-parameter dynamics \cite{onuki_6, binder1_6, bray_6, jones_6, lifs_6, amar_6, majum_6}, 
while significant progress has been made with respect 
to the understanding of structure and dynamics during phase separation in passive 
systems \cite{onuki_6, binder1_6, bray_6, jones_6, lifs_6, binder2_6, binder3_6, siggia_6, furu1_6, furu2_6, roy_6, royp_6, shimi_6, mid1_6},
in the subdomain of active matter systems 
 \cite{marche_6, rama1_6, cates1_6, mishra1_6, belmon_6, mishra2_6, cates2_6, hagan1_6, hagan2_6, wys_6, mehes_6,
peru_6, mishra3_6, cremer_6, schwarz_6, pala_6, kumar1, das1_6, tref_6, vic1_6, czi_6, bagl_6, chate_6, das2_6, schak_6, capri_6, loi_6},
though intense, the interest is 
rather recent. The basic questions that many of these studies ask are by concerning how the self-propulsion, a property inherent in the 
constituents of an active matter system, affects the universality classes \cite{onuki_6, binder1_6,bray_6,jones_6}
associated with such evolution dynamics. It is worth noting that there 
exist different types of self-propulsion. Influence of one type, on the structure and dynamics, is
expected to be different from the other. Primary objective of this work 
is to identify the effects of a certain type of self-propulsion, that
encourages the particles to align their directions of motion with each other \cite{vic1_6}, 
like in a system of inelastically colliding granular particles \cite{gold1}, on evolution during vapor-``solid'' transition
with a morphology that consists of well-separated clusters of ``solid'' phase, with short range order. 
\par 
~In the case of passive matter, following quenches of homogeneous 
configurations to state points inside the miscibility gap \cite{onuki_6, binder1_6}, as 
the evolution towards the phase-separated new 
equilibrium occurs, the average size ($\ell$) of particle-rich and particle-poor domains typically grows algebraically, with 
time ($t$), as \cite{onuki_6, binder1_6, bray_6, jones_6}
\begin{equation}
 \ell \sim t^{\alpha}.
\end{equation}
The patterns, formed by the above mentioned domains, usually exhibit the simple scaling property \cite{bray_6}
\begin{equation}\label{crl_scld}
 C(r,t) \equiv \tilde{C}(r/\ell),
\end{equation}
where $C(r,t)$ is a two-point equal time correlation function, defined as \cite{bray_6}
\begin{equation}\label{cofr}
 C(r,t)= \langle\psi(\vec{r},t) \psi(\vec{0},t)\rangle - \langle\psi(\vec{r},t)\rangle\langle\psi(\vec{0},t)\rangle.
\end{equation}
In Eq. (\ref{cofr}) $\psi$ is a space ($\vec{r}$) and time dependent order parameter field, 
a scalar quantity for the present problem, and $\tilde{C}$ is a time-independent master 
function.
 The scalar notation $r$, for the separation 
between two points in space, in the argument of $C$, is 
used with the understanding that there exists structural isotropy. The 
scaling property of Eq. (\ref{crl_scld}) implies that the growth is self-similar in nature \cite{bray_6}, 
i.e., apart from a change in 
the global length scale, the patterns at two different times are similar to each other, in a statistical sense. 
We repeat, in addition to this scaling property, in the 
passive situation, a good degree of understanding has been obtained \cite{onuki_6, binder1_6, bray_6, jones_6, yeung1} 
on the analytical forms of the 
correlation function and values of the growth exponent $\alpha$, based on the conservation of order parameter, 
transport mechanism, space dimension ($d$), overall density or composition of the system, etc.
\par 

~~For active matter systems there has been growing interest in recent times 
\cite{marche_6, rama1_6, cates1_6, mishra1_6, cates2_6, 
hagan1_6, hagan2_6, wys_6, mehes_6, peru_6, mishra3_6, cremer_6, chate_6, das2_6}
 in the understanding of the above mentioned aspects, 
viz., the scaling behavior of structure, corresponding analytical form 
 and the growth law.
 There exist interest in both aligning and nonaligning dynamic (or 
active) interactions, e.g., in systems containing 
 Vicsek-like \cite{vic1_6} active particles and active Brownian particles (ABP) \cite{tref_6}. 
Given that the universality classes associated with dynamics, particularly     
with  the evolution dynamics that is being
discussed here \cite{bray_6}, are not very robust, there exists the need for 
examining situations of different types. As mentioned above, already in the passive case
the universality is decided by a variety of factors.
In each of these cases, the influence of an
 activity can be different from the others, this being true for all
types of self-propulsion. Our objective here is to quantify how the  
 Vicsek \cite{vic1_6} activity alters the class associated with 
the kinetics of phase separation  in a system with low density of particles
that gives rise to disconnected cluster morphology \cite{roy_6, royp_6, mid2_6}.  For comparative purpose we have also presented some results from an ABP system.
\par

~~Many studies, with the above mentioned purpose, use models that do not exhibit phase transition in the passive limit. 
In such a situation, the effects of activity, in certain ways,  is less transparent. 
Following recent works \cite{das2_6,schak_6}, here we 
consider a model that has a passive limit which also undergoes phase transition. 
In these earlier works \cite{das2_6, schak_6} focus was on the understanding of pattern, growth and aging 
for high enough particle density so that the resulting vapor-liquid transition exhibits percolating or bicontinuous 
evolution morphology consisting of elongated high and low density regions of active particles. 
In contrast, here we undertake a comprehensive study to quantify the influence of Vicsek-like alignment activity on the
kinetics of 
vapor-``solid'' transition in $d=2$, for disconnected morphology, that is achievable when the density of particles 
is rather low. 
We report important results on both structure and growth. 
\par

~~The rest of the paper is organized as follows. In Section II we discuss the model and methods. Results are 
presented in Section III. Finally, Section IV concludes the paper with a brief summary and outlook.

\section{Model and Methods}
~~The passive interaction among the particles, in our study,
has been modeled via the potential \cite{frenkel_6, allen_6, han_6, mid2_6}
\begin{equation}
 u(r) = U(r) -U(r_c) - (r-r_c)\left(\frac{dU}{dr}\right)_{r=r_c},
\end{equation}
 where 
\begin{equation}
 U(r) = 4\epsilon \Big[\Big(\frac{\sigma}{r}\Big)^{12}-\Big(\frac{\sigma}{r}\Big)^{6}\Big],
\end{equation}
the standard Lennard-Jones (LJ) pair interaction energy, with 
$\epsilon$ and $\sigma$ being the strength and diameter of interaction, respectively.  
Here $r$ is the inter-particle 
distance and $r_c~(=2.5\sigma)$ is a cut-off radius only within which the particles interact. 
The phase diagram, in the temperature ($T$) -
density ($\rho$) plane, for the vapor-liquid transition that this passive
model exhibits,  has been estimated earlier, for $d=2$ as well \cite{mid2_6}. The obtained 
values for the critical temperature ($T_c$) and critical density ($\rho_c$)
in this dimension are $\simeq 0.41\epsilon/k_B$ and $\simeq 0.37$,
 respectively, 
$k_B$
being the Boltzmann constant and density being calculated as $N/V$ (in 
appropriate dimensionless unit, see below for clarification), for $N$
particles residing within a volume $V$.
\par
~~We have introduced the activity in the system by following a 
Vicsek-like \cite{vic1_6} rule, as previously mentioned. According to this rule, a particle tries to align its 
velocity along the average direction \cite{das2_6}, 
\begin{equation}
 \vec{D}_N = \frac{\sum_j \vec{v}_j}{\rvert \sum_j \vec{v}_j\rvert},
\end{equation}
of its neighbors contained within the circle of radius $r_c$,
with $\vec{v}_j$ being the velocity of the $j$th neighbor. As explained below, 
this interaction is implemented in our simulations in such a way that at each instant of time 
the particles will get only directional impact along respective $\vec{D}_N$, in addition to experiencing 
forces due to the passive interactions.
\par
~~We perform time-step driven molecular dynamics (MD) simulations \cite{frenkel_6, allen_6}
in $2D$ square boxes of linear dimension $L\sigma$, 
with periodic boundary conditions applied in both the directions. 
The dynamical equations are numerically integrated by using the Verlet velocity algorithm \cite{frenkel_6}.
To keep the temperature of the system constant, we have used the Langevin thermostat \cite{frenkel_6, allen_6}. 
Thus, for particle $i$, we have worked with the equation \cite{das1_6, tref_6, das2_6, frenkel_6}
(the dots imply time derivatives)
\begin{equation}\label{langv_eqn}
 m_i \ddot{\vec{r}}_i = - \nabla u_i -\gamma m \dot{\vec{r}}_i + \sqrt{(2\gamma k_B T m)}\vec{R}_i(t) + \vec{f}_i,
\end{equation}
where $m_i$ is the  mass (same for all the particles), $\vec{r}_i$ is the position, $\gamma$ is the damping coefficient, 
$u_i$ is the (passive)
potential
energy, $\vec{R}_i$ is a random noise, $\vec{f}_i$ is the active force 
\cite{vic1_6, das2_6} and $T$ is the temperature to which the system is quenched. The random noise $\vec{R}_i$ 
is Delta-correlated over space and time, i.e., $R_i^{\mu}$ and $R_j^{\nu}$, the $\mu$ and $\nu$ Cartesian
components, corresponding to the $i$th and $j$th particles, respectively, satisfy \cite{han_6}
\begin{equation}
 \big\langle R_i^{\mu}(t)R_j^{\nu}(t^\prime)\big\rangle = \delta_{\mu\nu} \delta_{ij}\delta(t-t^\prime),
\end{equation}
 where $t$ and $t^\prime$ stand for two different times. 
As already stated, the force $\vec{f}_i$ acts along $\vec{D}_N$, i.e., 
\begin{equation}\label{activity}
 \vec{f}_i \propto f_A \vec{D}_N,
\end{equation}
 $f_A$ being the strength of the alignment interaction \cite{das2_6}.
\par
~~Without the last term, evolution  of Eq. (\ref{langv_eqn}), 
via the Verlet velocity rule,
starting from time $t$, gives the velocity of the particles at time $t+\Delta t$ in the passive limit,
where $\Delta t$ is the time step of integration. In our simulations, 
we have used $\Delta t=0.01\tau$, $\tau~(=\sqrt{m\sigma^2/\epsilon})$ 
being the LJ unit of time. Following this, the particle velocities,
at each time step, were further updated by incorporating $\vec{f}_i$. 
Note that Vicsek activity \cite{vic1_6}, as stated above, is supposed to change only the direction of motion. This feature is 
appropriately taken care of here \cite{das2_6}. It is known that in some active matter systems, e.g., in the case of ABP systems the temperature of the active particles settles to a value higher than the assigned number \cite{loi_6,loi1_6}. Because of the nature of the active interaction in the Vicsek case \cite{vic1_6}, this possibility does not arise.
\par
~~For the sake of convenience, in the rest of the paper we set $m$, $\sigma$, $\epsilon$, $k_B$ and $\gamma$ to unity. 
All our results will be presented for $L=1024$.
The positions and velocities of all the particles, in the initial configurations, have been taken randomly,
that mimics high temperature homogeneous phase.
The evolution dynamics has been studied after quenching such configurations to a final
temperature $T$  below $T_c$, with overall 
density set at $\rho= 0.05$, close to the vapor branch of the coexistence curve. 
Final quantitative results are presented after averaging over runs with $50$ independent initial configurations.
Unless otherwise mentioned, all our results will be for $T=0.1$. We have considered several values of $f_A$, viz., $f_A=0, 0.5, 0.6$ and $1$,  $f_A=0$ being the passive limit. Most of the results for active case, however, are 
presented for $f_A=1$. Details on the ABP model have been provided later.
\par
~~For the calculation of various observables, e.g., the correlation function,
we map the off-lattice configurations to lattice ones \cite{roy_6, royp_6}. 
Each point on the lattice has been assigned a value of $\psi$, the order-parameter.   
It is either $+1$ or $-1$, depending upon whether the  
 local density, that can be calculated from the number of particles present within a small area around that point, 
is higher or lower than a cut-off value $\rho_{\rm{cut}}$. In this work, 
we choose $\rho_{\rm{cut}}=0.5$. Quantities like the average mass ($m$) of the clusters,
that will turn out to be important for the quantification of growth,
were calculated by appropriately identifying the 
boundaries of the clusters \cite{roy_6, royp_6, paul2, paul3}.

\section{Results}
~~We divide this section into two sub-sections.
Results for the pure passive case are presented in the first subsection. The active 
matter results are presented in the second one. Unless otherwise mentioned, active matter 
results will correspond to the Vicsek model.

\subsection{Passive case}
~~We remind the reader that our objective is to study the situation 
when domains of the high density phase do not percolate. So, we have chosen a 
low value of the overall density, viz., $\rho=0.05$, which is significantly shifted towards 
the vapor branch of the 
coexistence curve \cite{mid2_6}, for the chosen final temperatures. 
\par

\begin{figure}[h!]
\centering
\includegraphics*[width=0.36\textwidth]{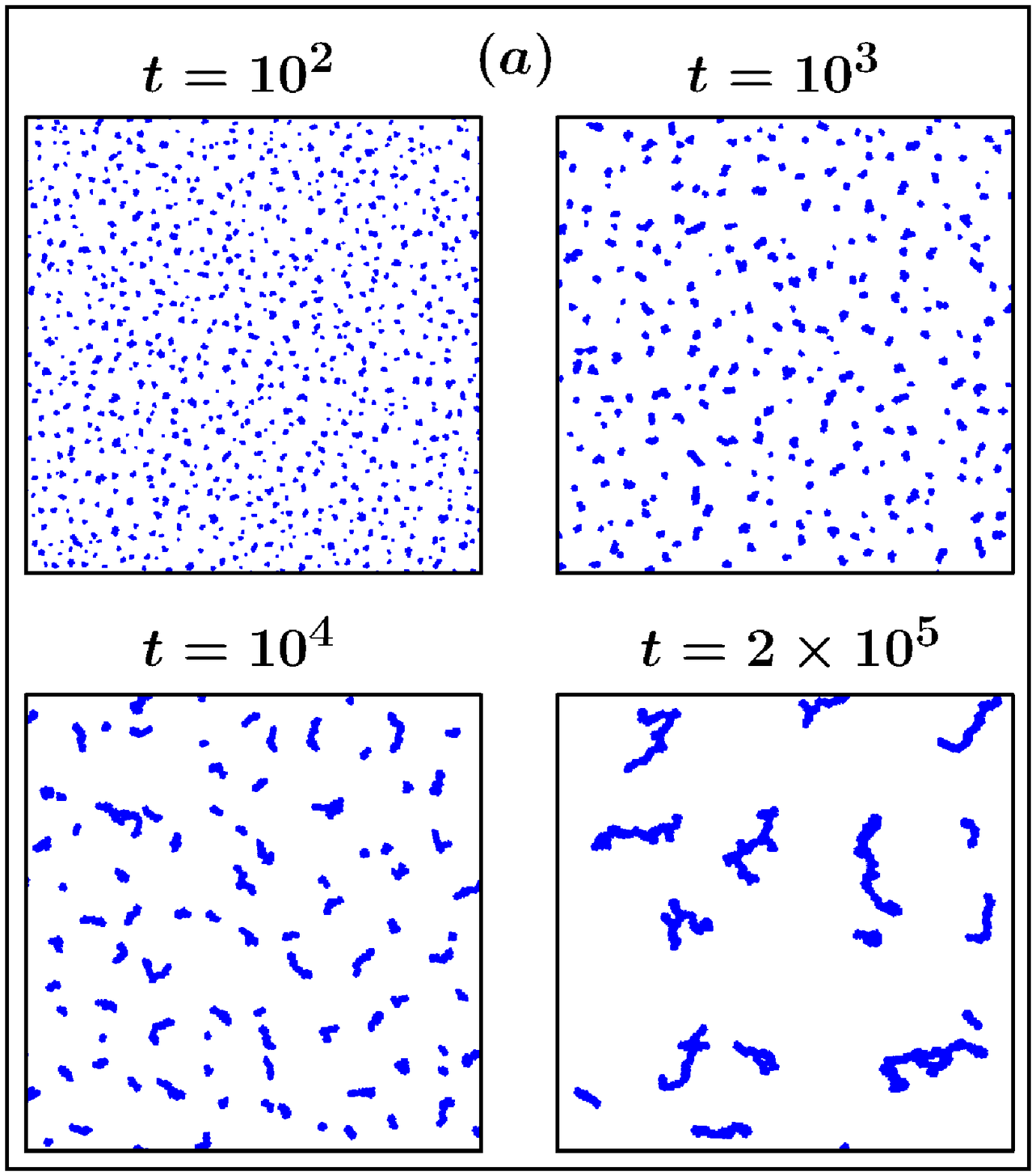}
\vskip 0.4cm
\includegraphics*[width=0.32\textwidth]{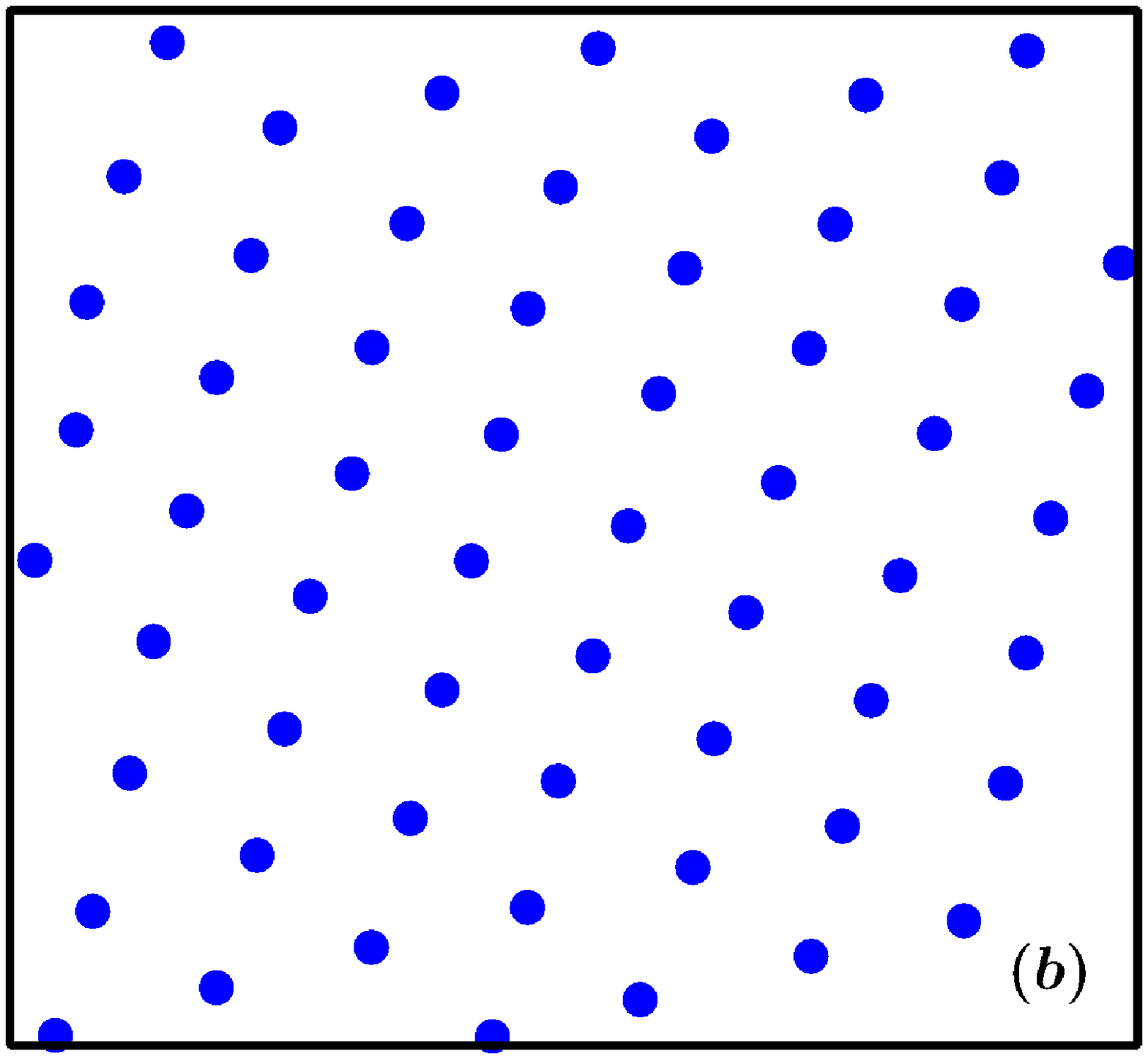}
\caption{\label{fig1_6} (a) Snapshots during the evolution of the model, 
in the passive limit, are shown from four different times, following quench of a 
random initial configuration to $T=0.1$. Locations of the particles are marked with dots.
(b) Part of a cluster from the snapshot at $t=2 \times 10^5$ in (a) is shown.}
\end{figure}

In Fig. \ref{fig1_6}(a) we show snapshots that are recorded during the evolution following the quench of a
random initial configuration  
to $T=0.1$. Cluster formation is quite evident even at $t=100$, the time for the 
earliest snapshot. One notices that at early enough time the droplets have  nearly circular appearance. With the 
progress of time, as the size of the droplets increases, the shape keeps deviating. 
The clusters at the latest presented time is very much filamental or fractal. 
We describe below a possible reason behind the formation of such fractal structure.

\begin{figure}[h!]
\centering
\includegraphics*[width=0.4\textwidth]{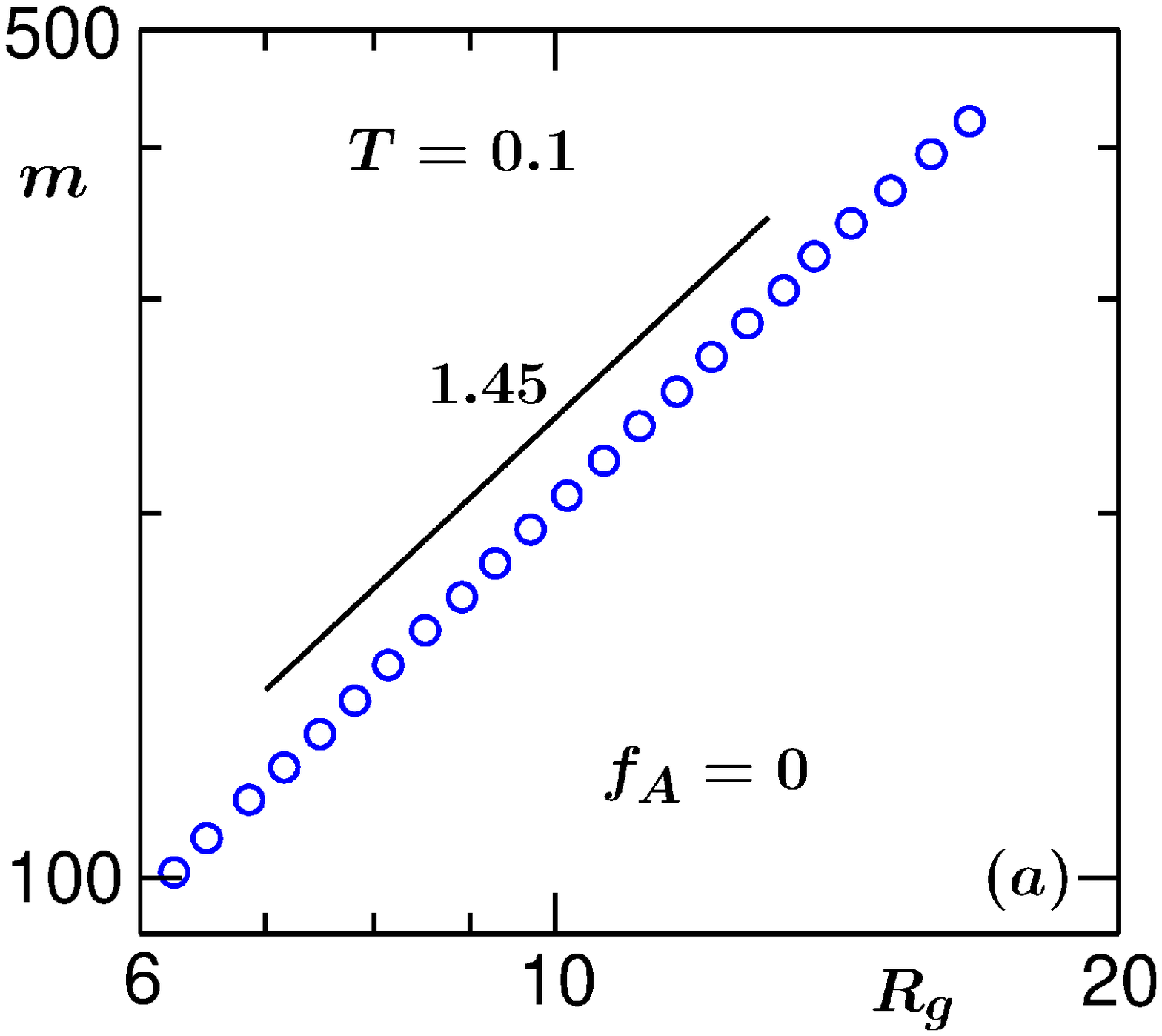}
\vskip 0.4cm
\includegraphics*[width=0.37\textwidth]{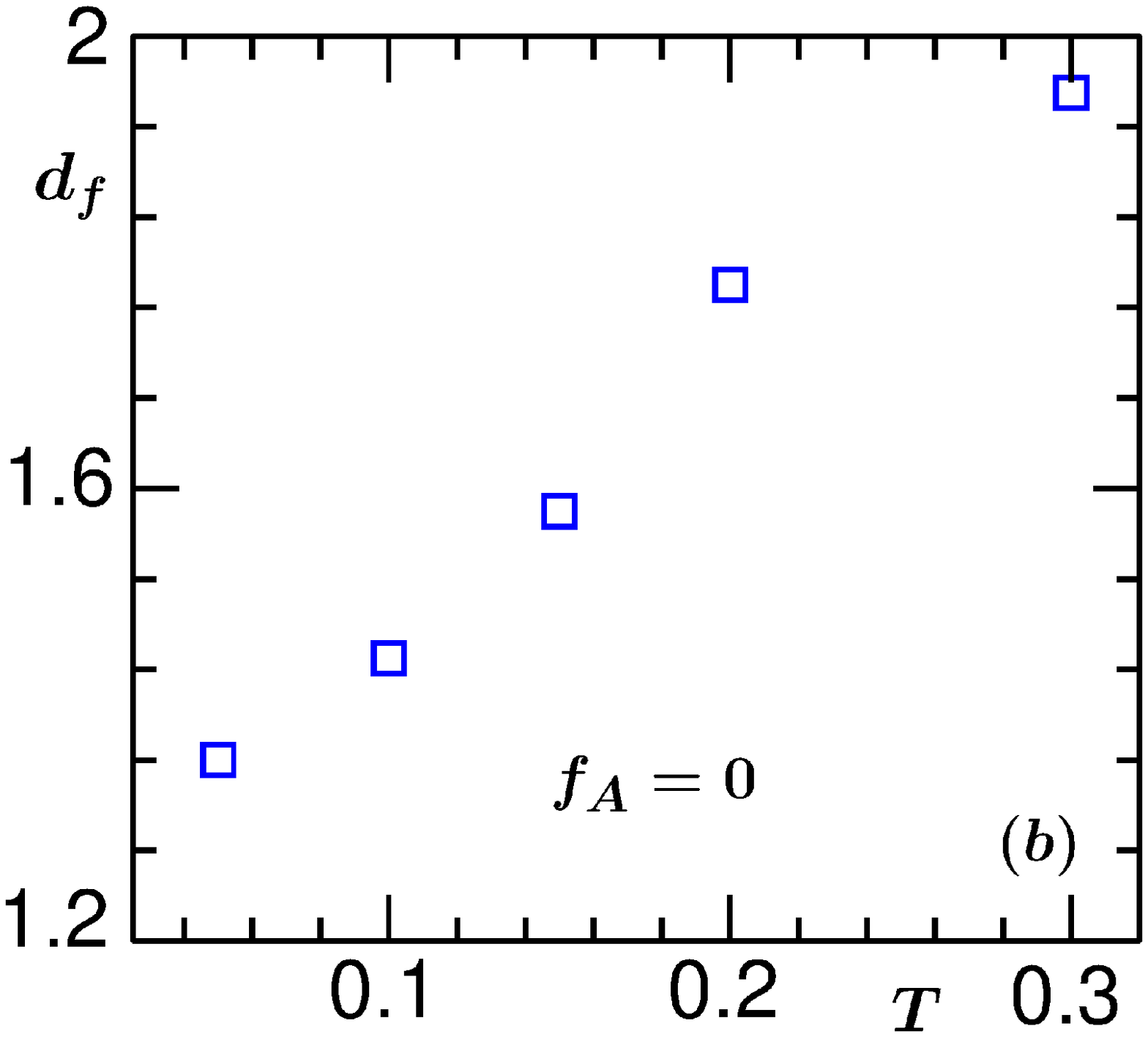}
\vskip 0.4cm
\includegraphics*[width=0.4\textwidth]{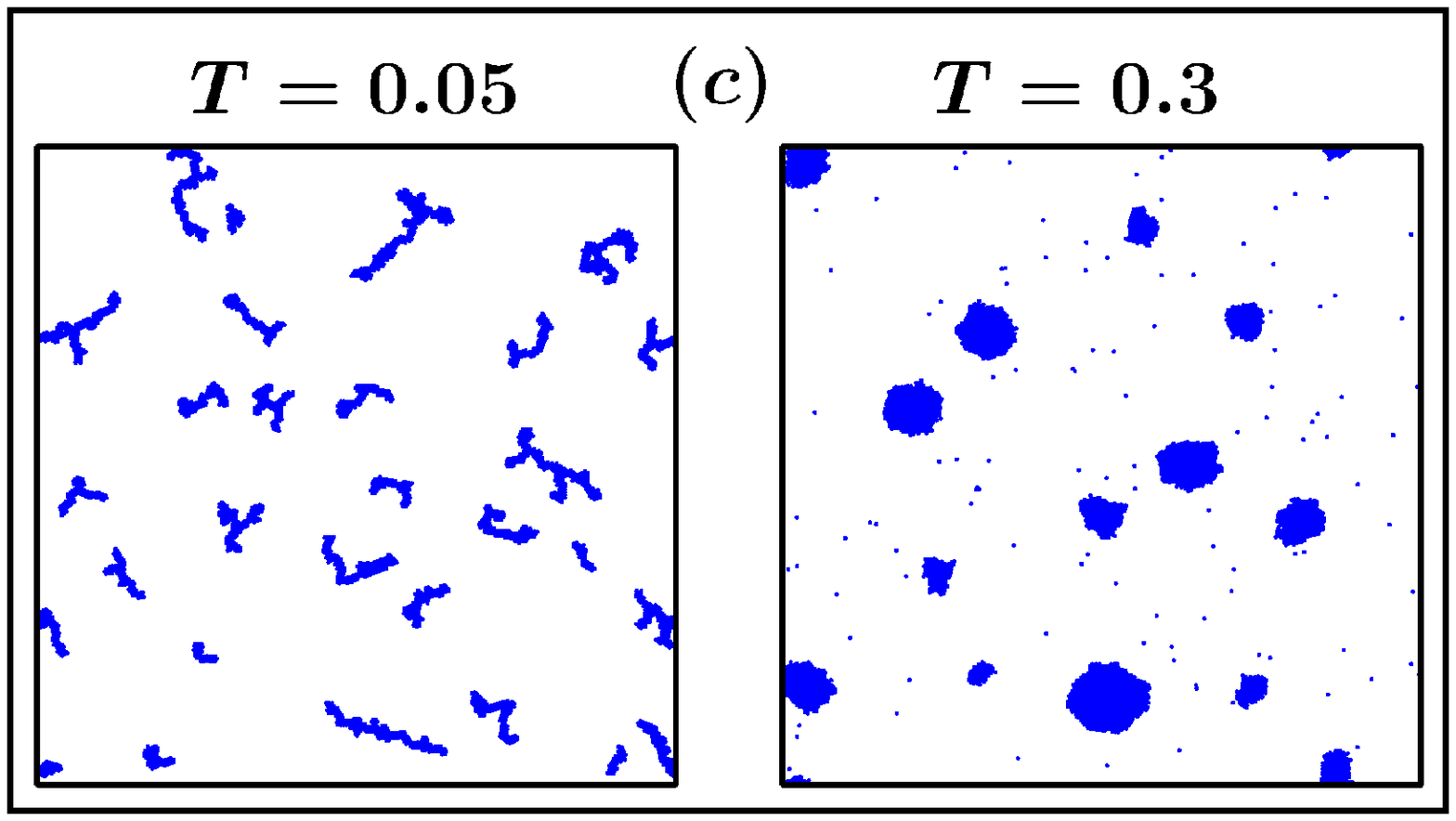}
\caption{\label{fig2_6} (a) Average mass of clusters ($m$) is plotted versus the 
average radius of gyration ($R_g$), on a log-log scale, for quenches of random
initial configurations to $T=0.1$. The solid 
line represents a power law, exponent for which is mentioned. 
(b) Fractal dimension, $d_f$, of clusters is plotted as a function of quenched temperature.
(c) Evolution snapshots are shown for two different temperatures, 
values of which are mentioned. In each of the cases we have chosen $t=10^5$. 
All results correspond to the passive limit of the model.}
\end{figure}

\begin{figure}[h!]
\centering
\includegraphics*[width=0.4\textwidth]{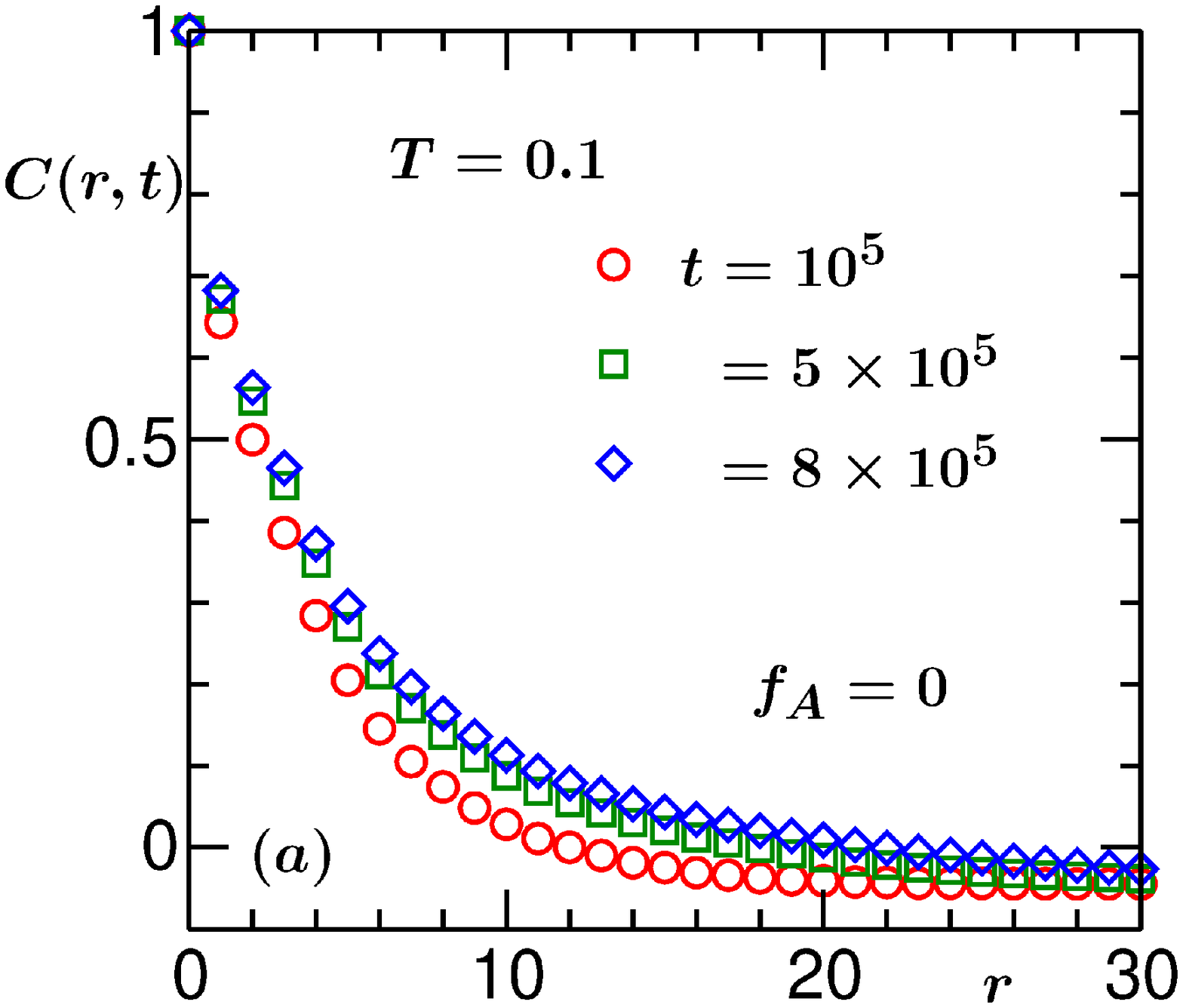}
\vskip 0.4cm
\includegraphics*[width=0.4\textwidth]{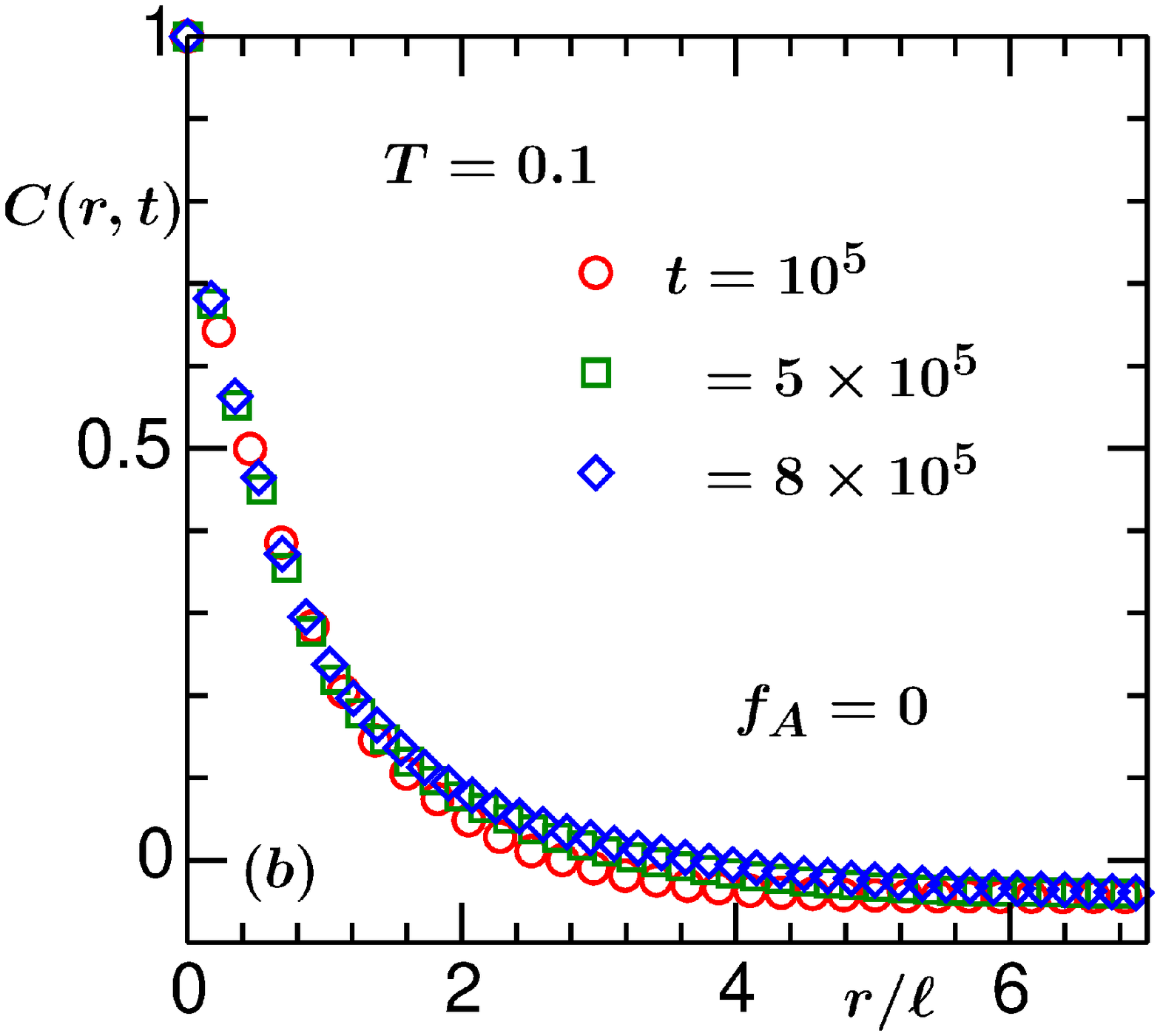}
\vskip 0.4cm
\includegraphics*[width=0.38\textwidth]{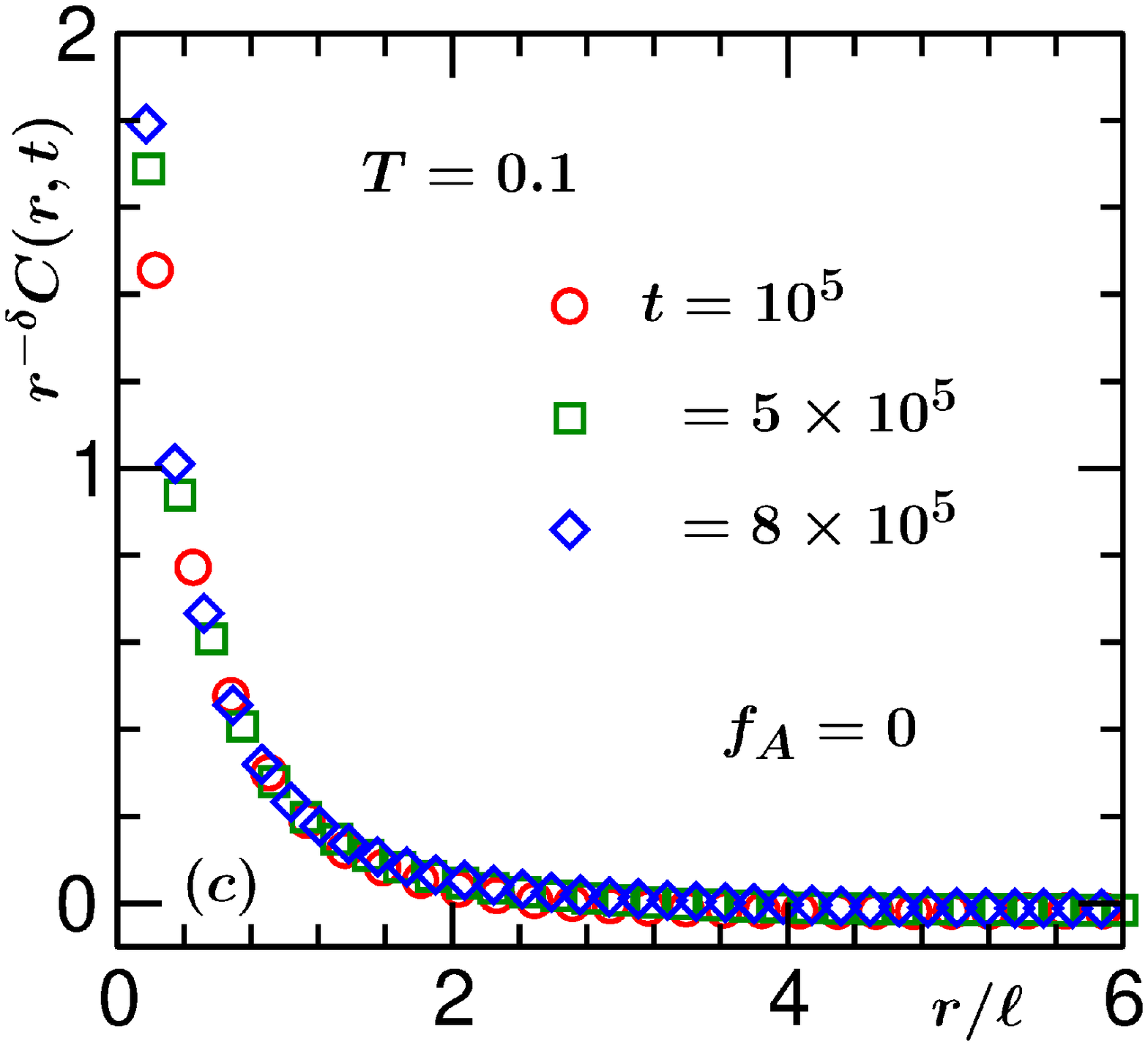}
\caption{\label{fig3_6} (a) Two-point equal time correlation functions, $C(r,t)$, are plotted versus $r$, the scalar distance between two space points. Data from three different times are 
shown. (b) $C(r,t)$ from different times are plotted versus the scaled distance $r/\ell$. (c) Scaling plots of $C(r,t)$ are shown after taking into account the correction 
factor due to the fractality of the structure (see text for details). 
All results are from the passive limit of the model, with $T=0.1$.}
\end{figure}

\begin{figure}[h!]
\centering
\includegraphics*[width=0.48\textwidth]{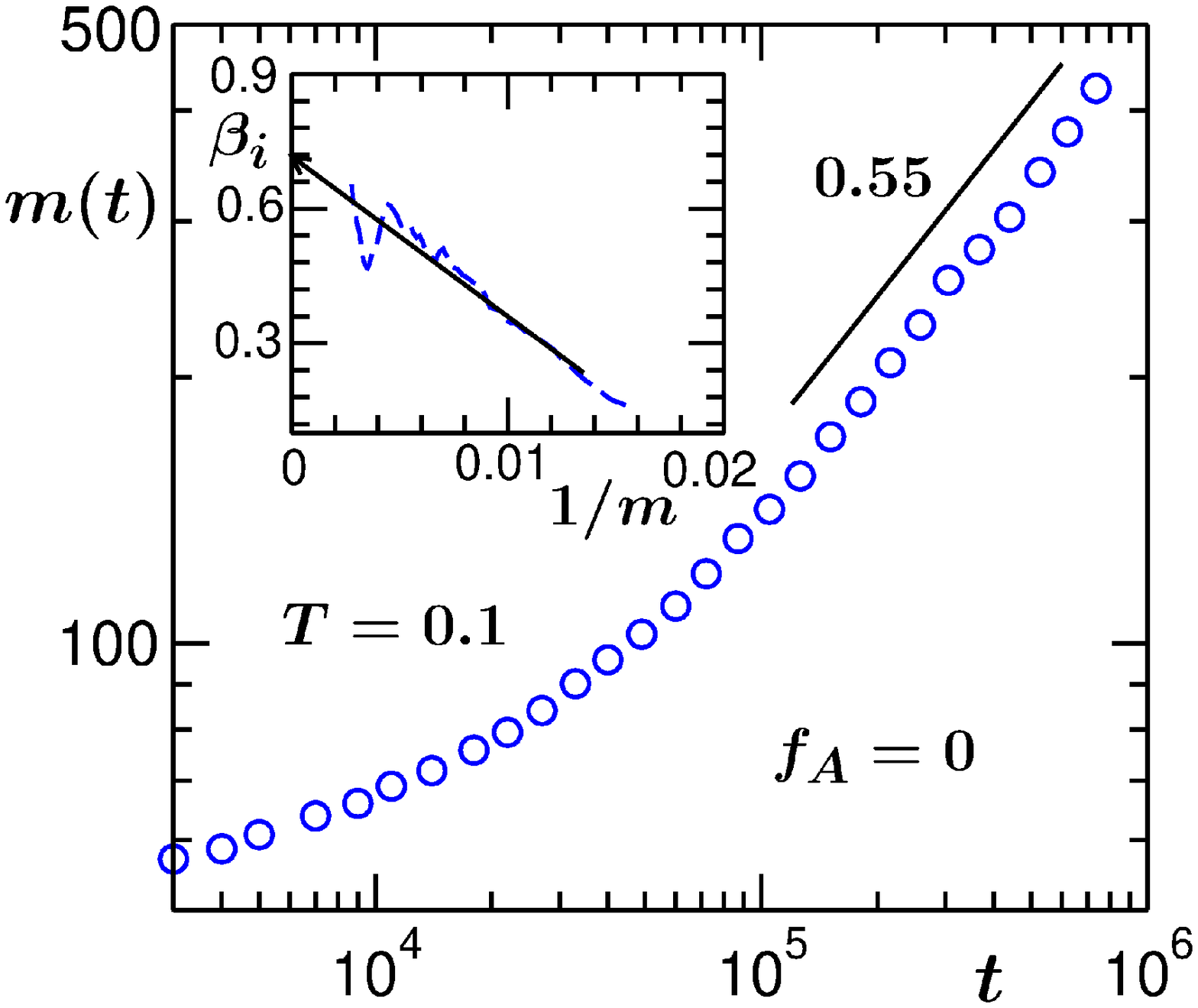}
\caption{\label{fig4_6} Average cluster mass ($m$) is plotted versus time, for the passive limit of the model. The solid line represents a power-law 
having exponent $0.55$. Inset shows $\beta_i$, the instantaneous exponent
(see text for the definition), as a function of $1/m$. The arrow-headed solid line there is a 
guide to the eye. These results are for quenches to $T=0.1$.}
\end{figure} 

\par
~~Because of the choice of a very low temperature, the clusters, i.e., the
the high density regions, are in a ``solid''-like state \cite{mid1_6}. 
These rigid clusters are essentially static, in translational sense, and 
growth occurs via the diffusive deposition \cite{lifs_6} of particles, on the larger
clusters, from the vapor phase, that are supplied by the smaller clusters. 
Of course, during this growth process 
the clusters try to obtain a 
circular shape, requirement for the minimization of interfacial free energy, 
via rearrangement of the particles. However, because of the solid-like arrangement of particles 
[see Fig. \ref{fig1_6}(b) where we show an enlarged view of a cluster from the snapshot at $t=2\times 10^5$] 
at very low temperature, this process  
(with relaxation time $\tau_1$) is not fast enough compared to the process of deposition of 
particles on the clusters (having relaxation time $\tau_2$). Therefore, once the structure deviates from the circular shape, 
quick regaining of the shape does not become 
possible. Furthermore, collisions of the particles, while being deposited from the vapor phase, 
with the clusters, can induce random rotations in the clusters. This makes the 
deposition more probable in anisotropic manner, resulting in well-grown filamental structures \cite{mid1_6}. 

\par
~~ To calculate the (mass) fractal dimension, in Fig. \ref{fig2_6}(a) we plot the average mass, $m$, of the clusters, as a function of the average radius 
of gyration, $R_g$. For the calculation of these quantities,
as previously stated, the cluster boundaries were 
appropriately identified \cite{roy_6, royp_6}. Number of particles in such a 
closed boundary provides the mass ($m_c$) of the corresponding 
cluster. The average value was obtained from the first moment of the related 
distribution. The radius of gyration of a cluster was estimated as \cite{gold_6}
\begin{equation}
 R_g^c = \Big[\frac{1}{m_c} \sum_{i=1}^{m_c} (\vec{r}_i - \vec{r}_{\rm{cm}})^2 \Big]^{1/2},
\end{equation}
where $\vec{r}_{\rm{cm}}$ is the location of the centre of mass of the cluster:
\begin{equation}
 \vec{r}_{\rm{cm}} = \frac{1}{m_c} \sum_{i=1}^{m_c} \vec{r_i}.
\end{equation}
Again, the average value was estimated from the first moment of the distribution of $R_g^c$. 
On a log-log scale, the plot in 
Fig. \ref{fig2_6}(a) has a linear appearance, in the large mass limit, i.e., in the long time regime. 
This implies a power-law behavior
\begin{equation}\label{frac_d}
 m \sim R_g^{d_f},
\end{equation}
$d_f$ being the fractal dimension \cite{vic2_6, vic3_6}. The data set appears consistent with the solid line that has the exponent $d_f=1.45$. Such 
small dimension was observed in Brownian dynamics simulations as well \cite{scior1_6, scior2_6}, for similar systems.

\par
~~If the fractal structure is a result of the competition 
between the time scales $\tau_1$ and $\tau_2$, we expect $d_f$ to have a 
temperature dependence. This is because, with the increase of the latter, $\tau_1$ decreases, whereas $\tau_2$ increases, because of 
decreasing cluster rigidity and increasing 
density of particles 
in the vapor phase, respectively. In Fig. \ref{fig2_6}(b) we plot $d_f$ as a function of $T$. Clearly, $d_{f}$ 
has a strong dependence on $T$, the former gets enhanced with the increase of the latter. For $T$ close to the triple 
point, which is around $0.3$ for this model, it appears, $d_f$ almost coincides with 
$d~(=2)$. For visual illustration, in Fig. \ref{fig2_6}(c) we have shown two typical 
snapshots, one from very low temperature and the other from $T=0.3$.
While extremely filament like structure is prominent at the lower temperature, 
all the clusters at $T=0.3$ have nearly circular shape. 
Rest of the results are presented from $T=0.1$, for both passive and active cases.

\par
~~In Fig. \ref{fig3_6}(a) we show plots of the two-point equal time correlation function,
$C(r,t)$, from different times, since the instant of quench, with the variation of distance $r$. Slower decay with 
increasing time implies growth in the system. To verify the scaling property of 
Eq. (\ref{crl_scld}), in Fig. \ref{fig3_6}(b) we show plots of $C(r,t)$ by dividing the 
distance axis by the ``average length'' ($\ell$)
of the domains. The latter is obtained as the distance at which $C(r,t)$ decays to 
$0.25$ times its maximum value that is, throughout the paper, normalized to unity. 
The data collapse at large values of $r/\ell$
does not appear good. This is because of the fractality. In such situations, appropriate scaling form is \cite{vic2_6, vic3_6}
\begin{equation}
 C(r,t) \equiv r^{\delta} \tilde{C}(r/\ell),
\end{equation}
where $\delta=d-d_f$. In Fig. \ref{fig3_6}(c) we have obtained excellent collapse of data by using the above form. 
The exercise in Fig. \ref{fig3_6}(c), in addition to validating the scaling form for fractal
structures, confirms that our estimation of $d_f$ is correct.

\par
~~To avoid the complexity of dealing with the fractal structures, it is 
instructive to examine the time dependence of average mass to probe the growth in such systems. 
Of course, one can calculate $R_g$ as a function of time,
which is the true characteristic length. Nevertheless, we adopt time dependence 
of $m$ as the marker. In Fig. \ref{fig4_6}, 
we have shown $m$ as a function of $t$, on a log-log scale. The data at late time tend to appear linear, 
implying power-law growth. Here we expect 
\begin{equation}
	m \sim t^{\beta},~~\mbox{with}~ \beta=\frac{2}{3}.
\end{equation}
This is because of the fact that the growth occurs via diffusive deposition of particles, referred to as the 
Lifshitz-Slyozov mechanism \cite{lifs_6, amar_6, majum_6, huse_6}.
 However, the exponent (see the number mentioned against the solid line that is 
consistent with the simulation data) appears significantly lower 
than this expected value. This is perhaps due to the fact that nucleation is delayed 
and there exists an off-set length or mass when the system enters the scaling regime. In such situations, instead of extracting the exponent 
from the log-log plots, one should adopt more accurate exercise. 
In the inset of this figure we plot the instantaneous exponent \cite{huse_6,paul2, paul3}
\begin{equation}
 \beta_i = \frac{d ({\rm{ln}}m)}{d ({\rm{ln}}t)}, 
\end{equation}
as a function of $1/m$, a standard practice in the literature for limited data span. 
One can appreciate that $\beta_i$, in our exercise, in the limit $m\rightarrow\infty$,
converges to $\beta \simeq 0.7$, 
very close to the expected value.

Next we investigate how the structure and growth, observed in the passive case, get
modified by the alignment interaction that is part of our general model. These findings and related
explanations are provided in the next subsection. Note that the final temperature remains $T=0.1$.

\subsection{Active case}

\begin{figure}[h!]
\centering
\includegraphics*[width=0.39\textwidth]{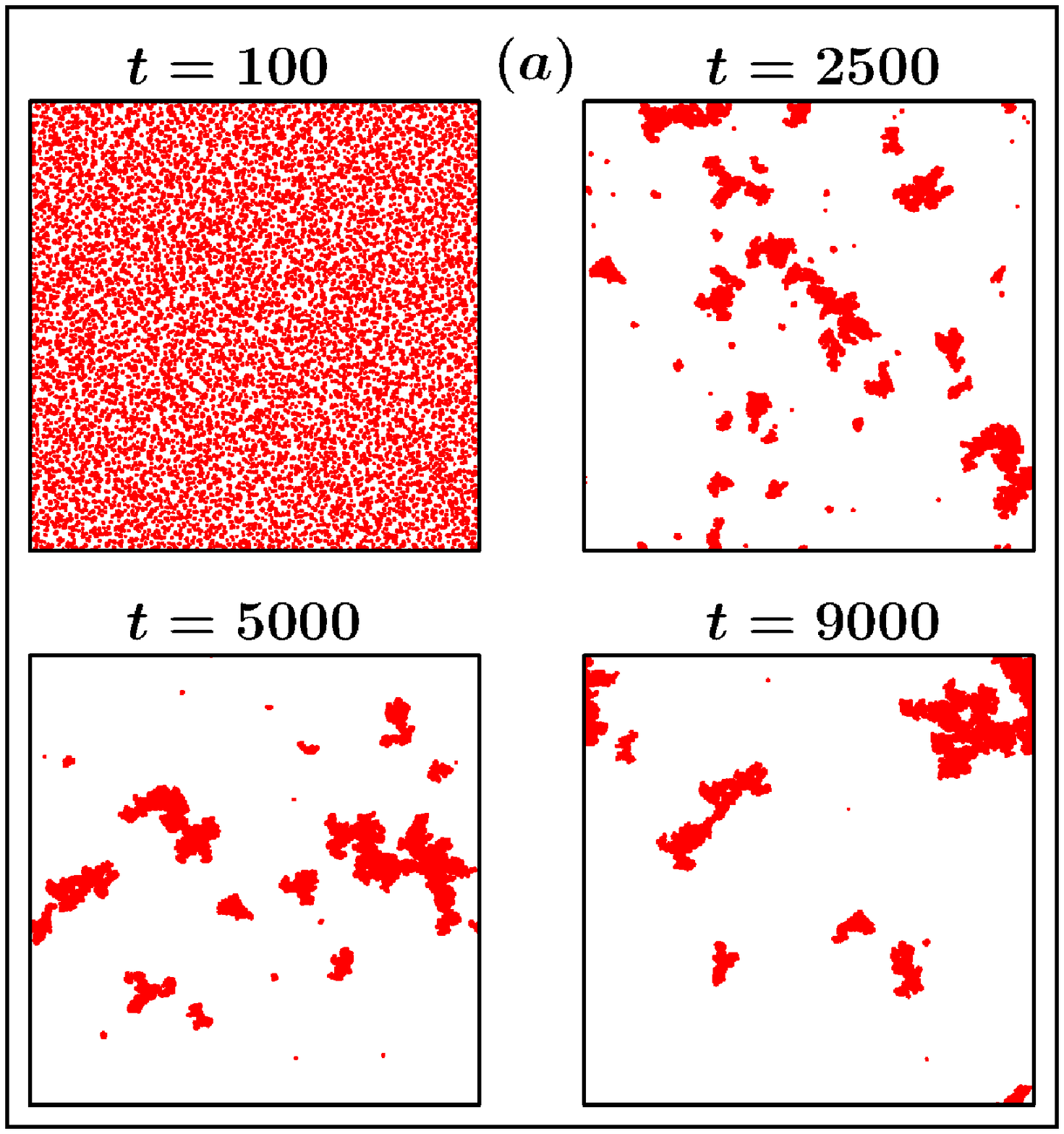}
\vskip 0.4cm
\includegraphics*[width=0.32\textwidth]{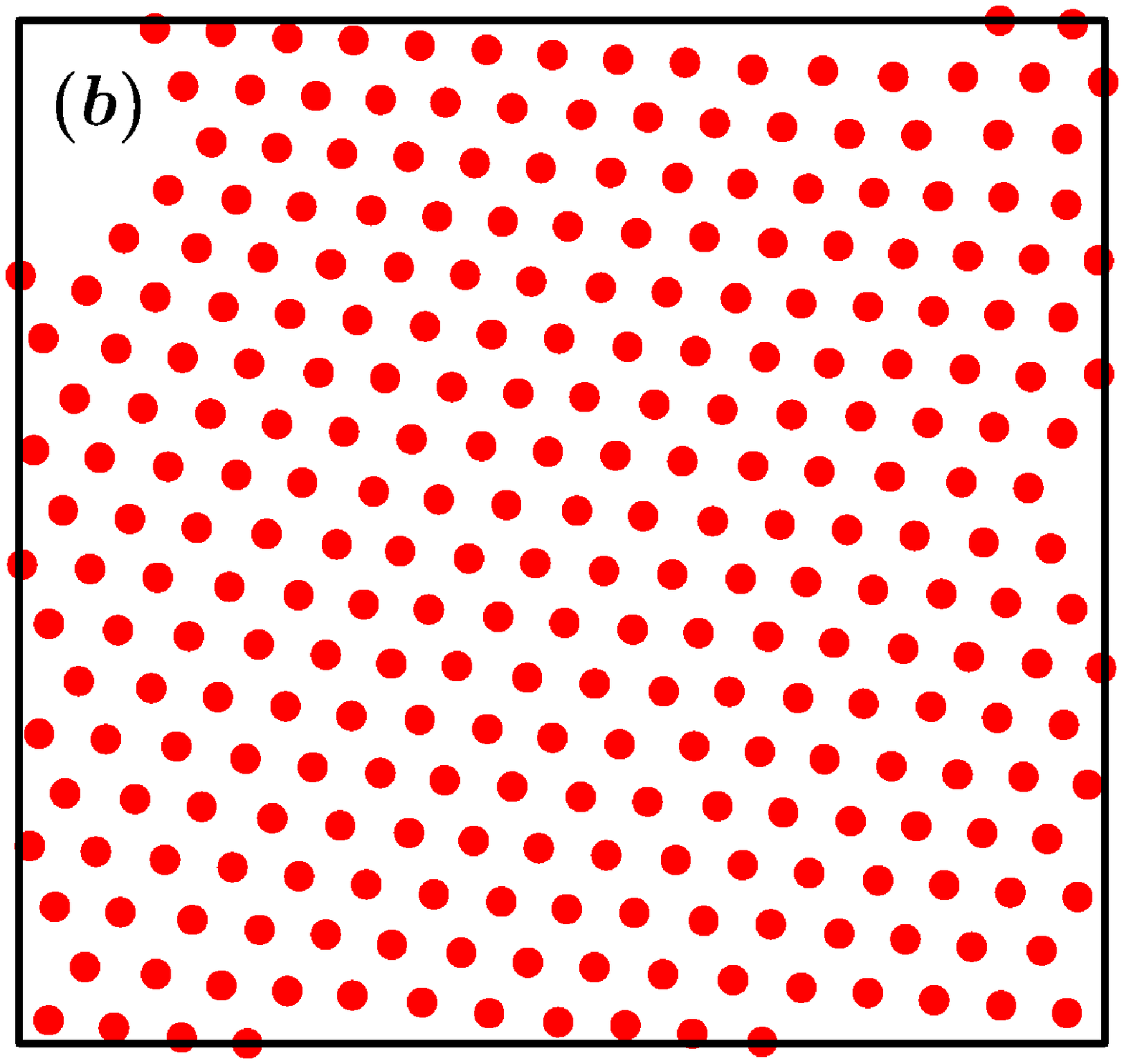}
\caption{\label{fig5_6} (a) Evolution snapshots, following quench of a random
initial configuration to $T=0.1$, are shown from four different times,
for the active case with $f_A=1$. (b) A portion of a cluster from the snapshot at $t=5000$ in (a) is shown. Note that the size of the portion here is larger than that in Fig. \ref{fig1_6}(b).}
\end{figure}

\begin{figure}[h!]
\centering
\includegraphics*[width=0.44\textwidth]{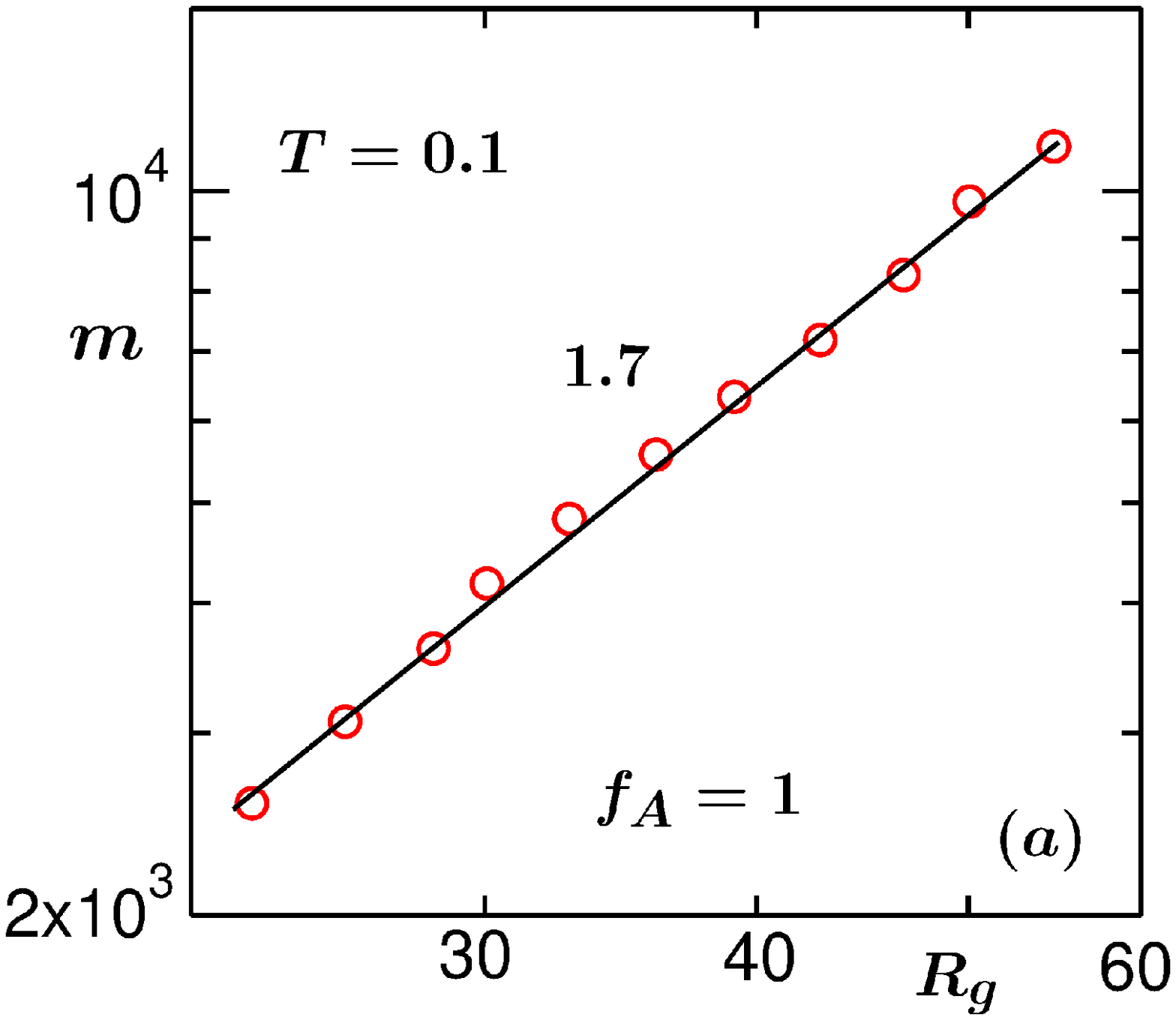}
\vskip 0.4cm
\includegraphics*[width=0.4\textwidth]{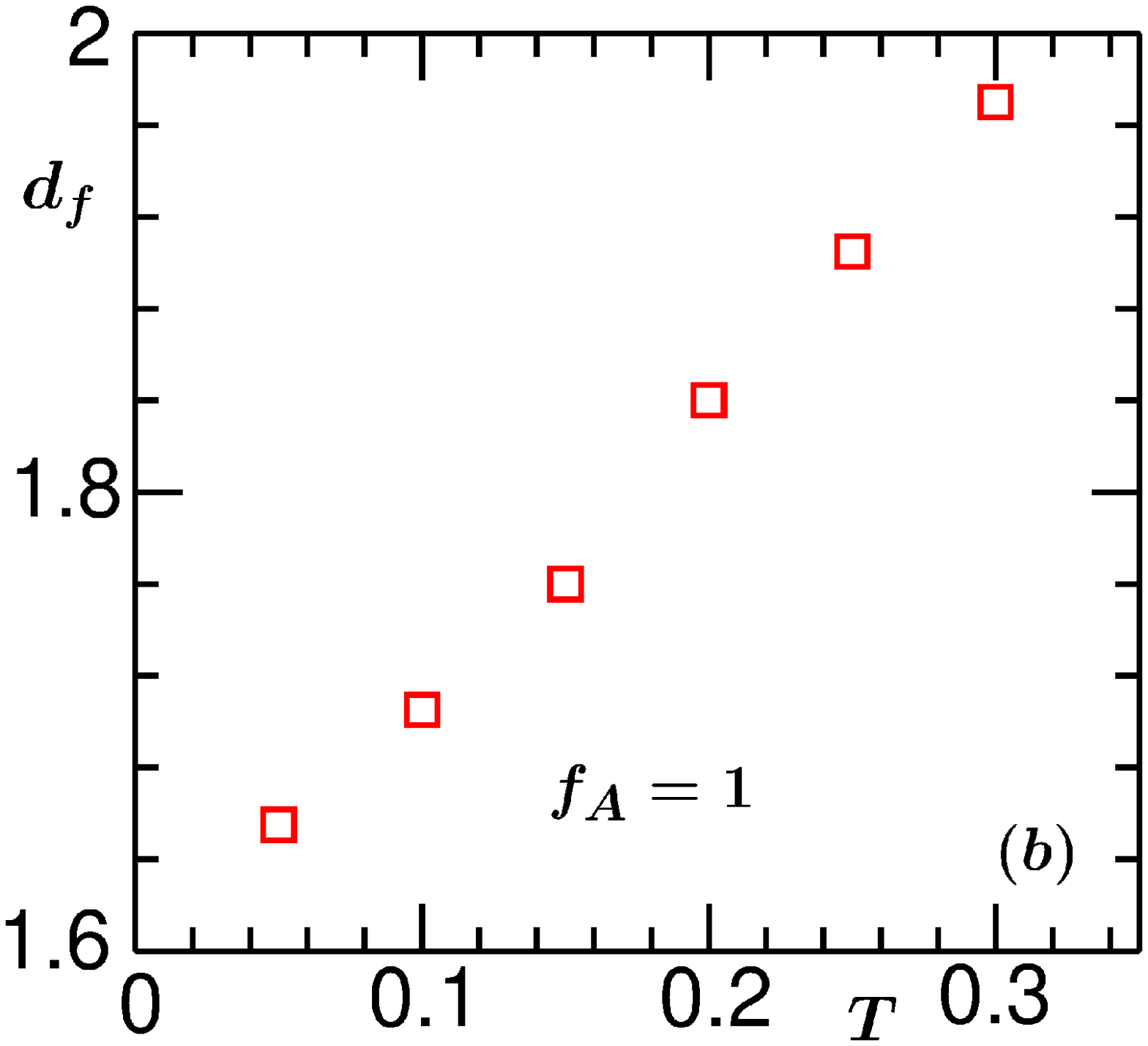}
\caption{\label{fig6_6} (a) Log-log plot of $m$ versus $R_g$, for $T=0.1$. The solid line is a power-law, 
representing the fractal dimension $d_f=1.7$. (b) Dependence of $d_f$ on $T$ is shown.
All results are for $f_A=1$.}
\end{figure}

~~As stated previously, unless specifically mentioned, all results in this subsection are for Vicsek activity and with $f_A=1$. In Fig. \ref{fig5_6}(a) we show snapshots from four different times, following quench
of a homogeneous configuration, for $f_A=1$. A comparison of these snapshots with those from Fig. \ref{fig1_6}(a) reveals that the 
fractality is lower in this case, i.e., $d_f$ is higher, 
even though the final temperature is the  same. 
A small part of a cluster from a late time snapshot is shown in Fig. \ref{fig5_6}(b). It appears that a ``solid''-like 
order exists even when the activity is turned on, at least for this value of $f_A$. As mentioned in the caption, the size of the part here is larger than that in Fig. \ref{fig1_6}(b). Because of this one may get the incorrect impression that the particle density is ``much'' higher inside the active clusters.
\par
At very  late time, when only a few clusters are left, there will be finite-size effects. In this limit the clusters are expected to assume circular shape, that is preferred for interfacial energy minimization, both for active and passive cases.
\par

To estimate the fractal dimension, in Fig. \ref{fig6_6}(a) we have shown a log-log plot of $m$ versus $R_g$,
like in the previous subsection. The linear 
look of the data set again indicates power-law behavior and 
the corresponding exponent provides $d_f \simeq 1.7$. Thus the structure is indeed less fractal than the passive case. 
 Such reduction in fractality can be due to the fact that the Vicsek 
\cite{vic1_6} activity 
keeps the particles within the clusters, at least the ones in the peripheral regions, 
mobile, with respect to the centres of mass. Thus, the time scale $\tau_1$ not being adequately smaller than $\tau_2$  
is of less relevance here. The temperature dependence of $d_f$, for $f_A=1$, is presented in Fig. \ref{fig6_6}(b).
Like in the passive case, here also $d_f$ tends to $d$ when $T$ approaches $0.3$. 

\par
\begin{figure}[h!]
\centering
\includegraphics*[width=0.45\textwidth]{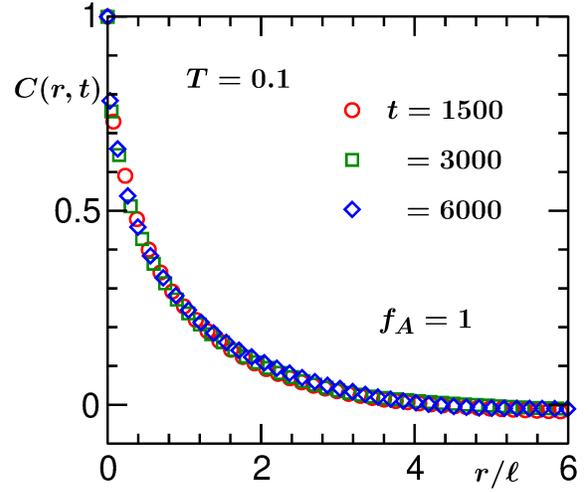}
\caption{\label{fig7_6} $C(r,t)$, from three different times, are plotted versus $r/\ell$, for the active case with $f_A=1$ and $T=0.1$.}
\end{figure}

~~In Fig. \ref{fig7_6} we show scaling exercise for $C(r,t)$, using data from three different times. 
The collapse appears reasonably 
good when the correlation functions are plotted versus $r/\ell$,
even without the introduction of $r^{\delta}$. The better quality of scaling 
when $C(r,t)$ is plotted versus $r/\ell$, 
compared to the passive case, is because of the higher fractal dimension,
that provides a small value of $\delta$. Next we move to quantify the growth.

\begin{figure}[h!]
\centering
\includegraphics*[width=0.48\textwidth]{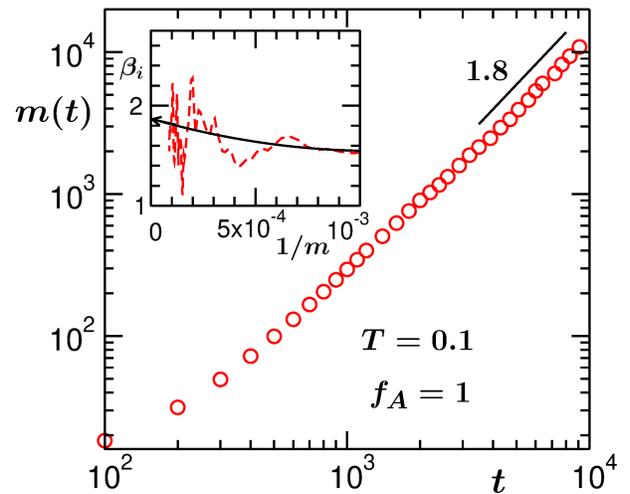}
\caption{\label{fig8_6} For $f_A=1$, the average mass of clusters is plotted versus time, on a log-log scale. The solid line is a power-law, 
exponent being mentioned next to it. Inset shows $\beta_i$ as a function of $1/m$. The arrow-headed solid line there is a guide to the eye.
All results are for $T=0.1$.}
\end{figure}

\par
~~Unlike the passive case, here the clusters can move, because of the activity. 
This may lead to growth primarily via certain cluster coalescence mechanism. For 
diffusive motion of the clusters, Binder and Stauffer \cite{binder2_6, binder3_6, siggia_6} 
pointed out that the growth exponent $\beta$ should be $1$.
This value of the exponent emerges  from the solution of the equation \cite{siggia_6}
\begin{equation}
 \frac{dn}{dt} = -Cn^2,
\end{equation}
where $n$ is the cluster density ($\varpropto 1/m$) and $C$ is a constant, 
a consequence of the Stokes-Einstein-Sutherland relation \cite{onuki_6,han_6,sengers}. 

\par
~~In Fig. \ref{fig8_6} we present a plot of $m$ as a function of time,
for the present problem. On the log-log scale the data set appears consistent with a power-law,
at least in the long-time limit. 
However, the exponent is much higher than unity (see also the plot of instantaneous exponent 
$\beta_i$ versus $1/m$, in the inset, for being better convinced), 
that was expected for diffusive coalescence mechanism \cite{binder2_6, binder3_6, siggia_6}. 
A reason behind such a sharp 
disagreement can be the presence of fractal feature in the structure, 
along with a possibility that the motion of the droplets in this case is much faster than
simple diffusion. To investigate the latter we calculate 
the mean-squared-displacement (${\rm{MSD_{CM}}}$) of the centres of mass of the 
clusters \cite{han_6}. 

\par
\begin{figure}[h!]
\centering
\includegraphics*[width=0.45\textwidth]{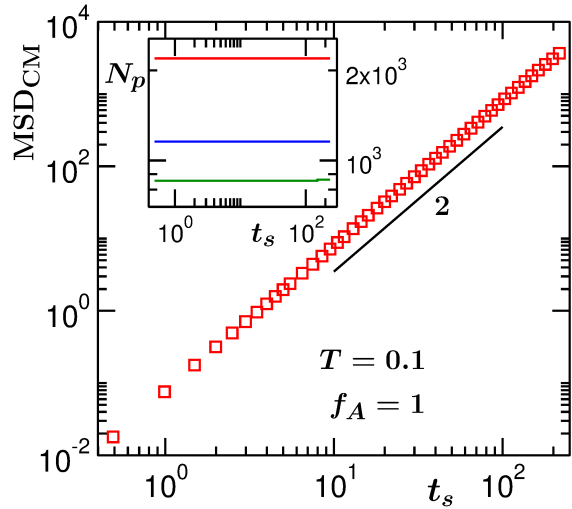}
\caption{\label{fig9_6} Mean-squared-displacement of the centre of mass of a cluster (${\rm{MSD_{CM}}}$), for $f_A=1$, is plotted as a function of time, 
on a log-log scale. The solid line is proportional to $t_s^2$. The inset shows the numbers 
of particles ($N_p$) in a few different clusters, versus translated times, during periods within which they do
not undergo collisions with other clusters. All results are for $T=0.1$.}
\end{figure}

~~In Fig. \ref{fig9_6} we show a log-log plot of ${\rm{MSD_{CM}}}$, 
as a function of time, the latter being shifted with respect to a starting value, for a typical cluster. The data exhibit practically a quadratic 
behavior, implying growth due to ballistic-like aggregation mechanism \cite{carne_6, trizac1_6, trizac2_6},
rather than the diffusive coalescence mechanism \cite{siggia_6}. 
In the inset of this figure we also show the numbers of particles in a few clusters, with the progress of time. 
Practically flat behavior of these plots 
rules out the possibility of any significant contribution due to the  Lifshitz-Slyozov particle diffusion mechanism.

\par
~~Coming back to the plot of $\rm{MSD_{CM}}$ in the main frame of Fig. \ref{fig9_6}, as already mentioned, the time here is translated with respect to certain reference time and thus, is not related to the simulation time. One must note that such a plot here is meaningful only for the duration over which a cluster is not merging with another, following a collision. Thus the time scale of such a plot cannot match that of overall simulation.

\begin{figure}[h!]
\centering
\includegraphics*[width=0.44\textwidth]{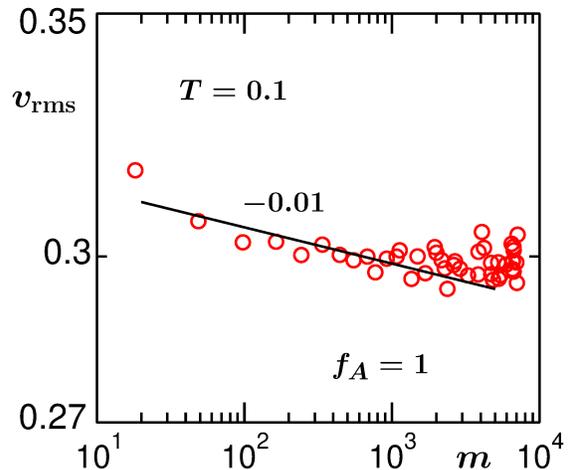}
\caption{\label{fig10_6} Log-log plot of the root-mean-squared 
velocity ($v_{\rm{rms}}$) of the clusters as a function of average mass, for $f_A=1$ and $T=0.1$. The solid 
line represents a power-law. }
\end{figure}

\par
~~Below we consider a theory of ballistic aggregation 
\cite{cremer_6, carne_6, trizac1_6, trizac2_6}
to see if the high value of the growth exponent, viz., $\beta \simeq 2$,
can be explained. 
For that purpose we write the kinetic equation \cite{mid1_6,carne_6,trizac1_6,trizac2_6,paul2,paul3}
\begin{equation}\label{n_evol}
 \frac{dn}{dt} = - ~{\mbox{``collision-cross-section''}} \times v_{\rm{rms}} \times n^2,
\end{equation}
where $v_{\rm{rms}}$ is the root-mean-squared velocity of the clusters. In $d=2$, the ``collision-cross-section'' is the 
radius of gyration which has the mass dependence $R_g \sim m^{1/d_f}$ [see Eq. (\ref{frac_d})]. Using this, 
and taking $n \propto 1/m$ and $v_{\rm{rms}} \sim m^{-z}$, in Eq. (\ref{n_evol}), 
one arrives at 
\begin{equation}\label{mass_evol}
 \frac{dm}{dt} = m^{(1-zd_f)/d_f}.
\end{equation}
Solution of Eq. (\ref{mass_evol}) provides  \cite{trizac2_6,paul3}
\begin{equation}
 m \sim t^{\beta}, ~~\mbox{with}~ \beta=\frac{d_f}{d_f(z+1)-1}.
\end{equation}
\par
~~It is worth mentioning that for such a picture to be valid, velocity relaxation, following the disturbance after a collision, should be much faster than the typical collision interval. This we have confirmed to be true via the calculations of time dependent average velocity within a newly formed cluster. It appears that a constant value  is rapidly reached, in fact, within a few MD steps.
\par

~~Given that $d_f \simeq 1.7$ and $\beta \simeq 2$, one should have $z \simeq 0.1$, 
a value much smaller than $0.5$, that is expected in situations when the velocities of the clusters
are random, typically observed in passive matter systems
\cite{mid1_6,carne_6,trizac1_6,trizac2_6,paul3}.
In Fig. \ref{fig10_6} we show $v_{\rm{rms}}$ as a function of $m$, for the present
system, i.e., for $f_A=1$. Indeed, 
the value appears much smaller than $0.5$. The discrepancy that is observed 
with the expectation, i.e., $z=0.1$, can be due to the following reason. It is possible 
that on an average ${\rm{MSD_{CM}}}$ 
deviates from the quadratic time dependence, to some extent. 
To ascertain that, of course, more accurate study, with very good statistics, 
is needed. However, our results are already in a very good agreement
with the theoretical picture, even at a quantitative level. 
Here note that, since in an active matter 
system energy is continuously injected to each particle, from the environment,
it is not surprising that $v_{\rm{rms}}$ will be nearly independent of mass. 
\par
\begin{figure}[h!]
	\centering
	\includegraphics*[width=0.45\textwidth]{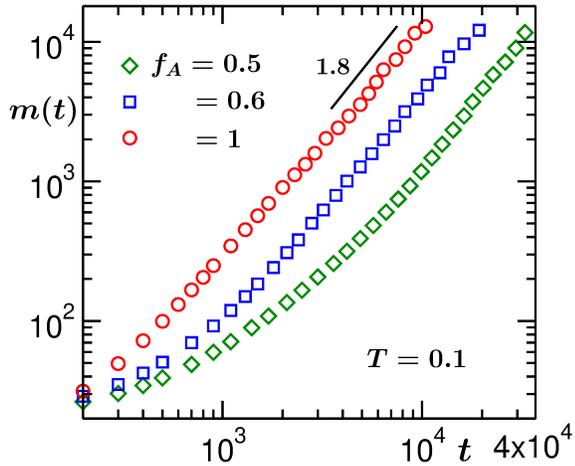}
	\caption{\label{fig11_6} Log-log plots of the average mass $m(t)$ of the clusters as a function of time, for $f_A=0.5, 0.6$ and $1$, with $T=0.1$. The solid	line represents a power-law. }
\end{figure}

~~In Fig. \ref{fig11_6} we present growth data from a set of different non-zero values of $f_A$. It appears, in each of the cases the asymptotic exponent is similar. However, the onset of the latter gets delayed with the decrease of $f_A$. Of course, there may be minor variation in the exponent as well, e.g.,  due to change in $d_f$. More systematic study is needed to uncover this.
\par
\begin{figure}[h!]
	\centering
	\includegraphics*[width=0.38\textwidth]{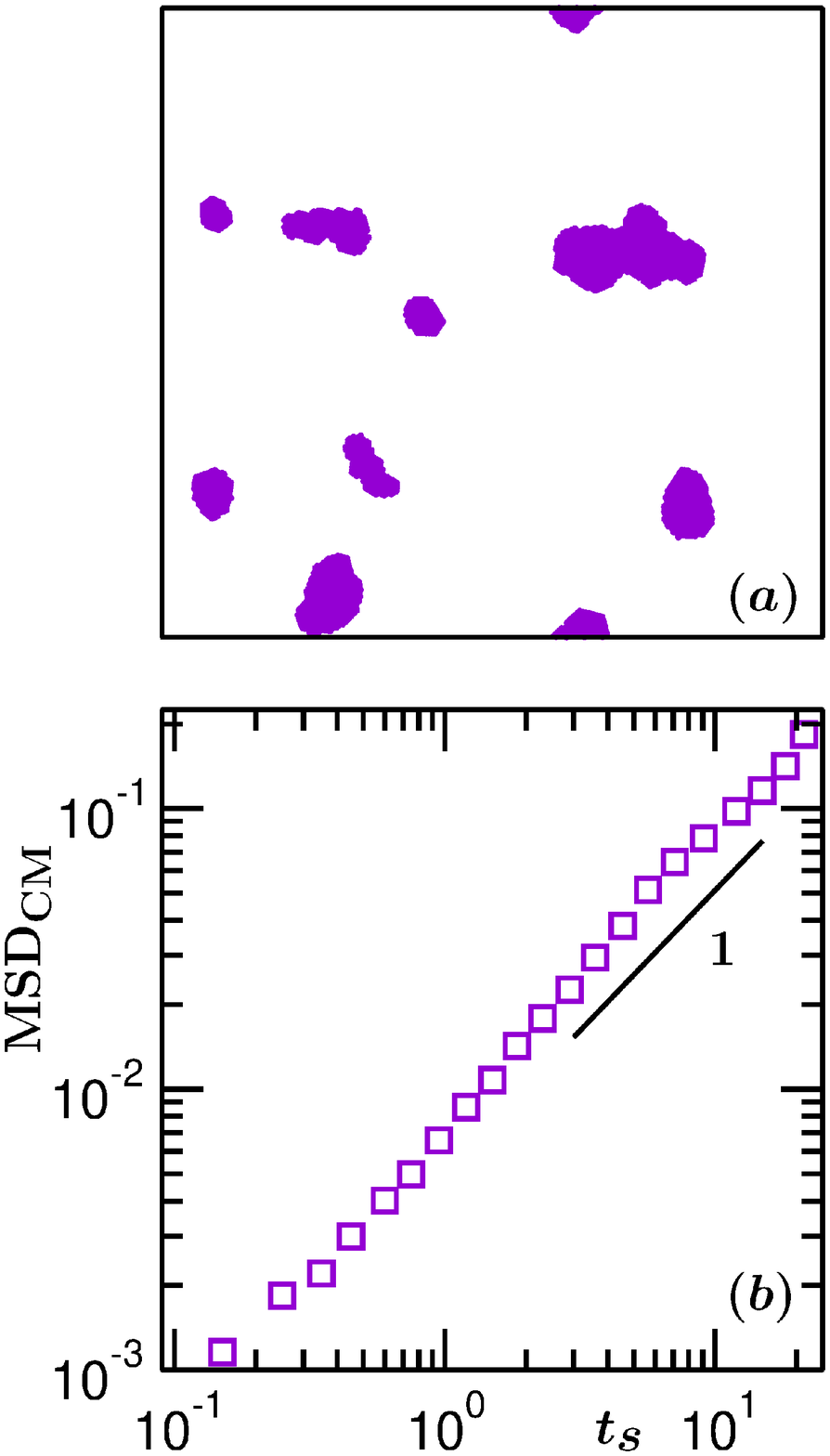}
	\caption{\label{fig12_6} (a) A typical evolution snapshot for a system containing active Brownian particles, with $f_p=1$ and background temperature $T=0.1$,  from $t=10^5$. (b) Depiction of time dependence of mean squared displacement of a cluster.}
\end{figure}

~~To further elucidate on the role of alignment interaction in the Vicsek model, in Fig. \ref{fig12_6} we show results for an ABP system. For this model the dynamical equations are \cite{sten_6}
\begin{equation}\label{trans}
 \dot{\vec{r}}_i = \beta D_t[-\nabla u_i+f_p \vec{p}_i]+\sqrt{2D_t}\Lambda_i^t,\\
 \end{equation}
 and
\begin{equation}\label{rot}
 \dot{\theta_i} = \sqrt{2D_r}\Lambda_i^r.
\end{equation}
Here $\vec{r}_i$ is the position of the $i$th particle, $u_i$ is the passive LJ potential as before, and $f_p$ is the strength of the self-propulsion force, having direction $\vec{p}_i\equiv (\cos \theta_i, \sin \theta_i)$, whereas $D_t$ and $D_r$ are, respectively, the translational and rotational diffusion constants of the particles. Furthermore, in Eq. (\ref{trans}) and (\ref{rot}) $\Lambda_i^t$ and $\Lambda_i^r$ are the zero-mean and unit-variance Delta-correlated random noises. Note that for the passive Brownian particles $f_p=0$. Here we have chosen $f_p=1$. For this simulation, we have set the integration time step at $10^{-3} \tau$.
\par 
~~In part (a) of Fig. \ref{fig12_6} we show a typical evolution snapshot. Clearly the phase separation is occurring \cite{capri_6}. However, the clusters in this case, as opposed to the Vicsek results,  do not move ballistically. This is shown in Fig. \ref{fig12_6}(b).

\section{Conclusion}

~~From extensive molecular dynamics simulations we have presented results 
for the kinetics of phase separation in passive and active 
matter systems. The passive system is the limiting case of a general active 
matter model. In a certain sense this set up is advantageous for the quantification of the effects of activity. 
The inter-particle interaction in the passive model is described by
a variant of the Lennard-Jones potential \cite{frenkel_6, allen_6, han_6}. 
The self-propulsion, on the other hand, for the main study on active case, is introduced via the well-known 
Vicsek model \cite{vic1_6} that facilitates phase separation via cooperative motion. 
The overall density of particles for our studies was chosen in such a way that the
morphology consisted of disconnected clusters.
\par
~~In our molecular dynamics simulations, the temperature was controlled via a Langevin thermostat \cite{frenkel_6}. 
In the passive limit this arrangement provides growth of solid state
clusters via particle diffusion mechanism \cite{lifs_6}, activated by concentration gradient. 
In the active case, on the other hand, we have identified that
the clusters grow practically via the ballistic aggregation mechanism \cite{mid1_6, carne_6, trizac1_6, trizac2_6}.
Exponent for the latter appears much higher compared to 
the Lifshitz-Slyozov \cite{lifs_6} value, outcome of the above mentioned diffusive mechanism.
The estimated growth law for the active case 
we have tried to explain by incorporating the information associated with the
fractality and velocity of the clusters in a relevant theory of
ballistic aggregation \cite{carne_6, trizac1_6, trizac2_6, paul2, paul3}. For this purpose, identification of ballistic motion and confirmation of negligible roles of the other mechanisms were important. We have also shown that in a system of active Brownian particles such motion of clusters is non-existent.
\par
~~It is expected that hydrodynamics \cite{frenkel_6, han_6} will play important
role in the growth process.  Thus,
it will be interesting to perform similar studies with a set up where
active particles are immersed in a hydrodynamic solvent.
Furthermore, the effects of fractality on the growth
can be checked in details by varying the final temperature ($T$)
and strength of activity ($f_A$). In this work we have already shown that
the fractality of the solid clusters increases with the decrease of 
both $T$ and $f_A$. Full scale studies of the kinetics with
the variation of these parameters will, thus, be useful. However,
such studies will bring additional complexity. The mean-squared-displacement
of the clusters will have different power-law time-dependence for different combinations of $T$ and $f_A$.
Thus, the outcomes will be difficult to interpret. Nevertheless, this
will provide a nice platform for the general understanding of cluster growth via
coalescence mechanism. We intend to pursue such avenue in future.
Furthermore, at a given temperature the effects of $f_A$ may be nonmonotonic \cite{mani_6}. To verify this one needs to consider very large range of $f_A$. This also we leave for a future study.

~~~~~
~~~~~

~~{\bf Acknowledgment:} 
SKD acknowledges financial support from the Marie Curie Actions Plan of European Commission 
(FP7-PEOPLE-2013-IRSES grant No. 612707, DIONICOS); Department of Biotechnology, India (grant No. LSRET-JNC/SKD/4539); 
and Science and Engineering Research Board of Department of Science and Technology, India (grant No. MTR/2019/001585).
SP is thankful to UGC, India, for research fellowship.

~~{\bf Conflicts of Interest}: There are no conflicts of interest to declare.

~$^{*}$ das@jncasr.ac.in


\begin{thebibliography}{100}
	\bibitem{onuki_6} A. Onuki, \textit{Phase Transition Dynamics}, Cambridge University Press, Cambridge, England, 2002.
	\bibitem{binder1_6} K. Binder, in \textit{Phase Transformation of Materials}, edited by R.W. Cahn, P. Haasen, and E.J. Kramer, VCH, Weinheim, 1991, p. 405, Vol. 5.
	\bibitem{bray_6} A.J. Bray, \textit{Adv. Phys.}, 2002, \textbf{51}, 481.
	\bibitem{jones_6} R.A.L. Jones, \textit{Soft Condensed Matter}, Oxford University press, Oxford, 2008.
	\bibitem{lifs_6} I.M. Lifshitz and V.V. Slyozov, \textit{J. Phys. Chem. Solids}, 1961, \textbf{19}, 35.
		\bibitem{amar_6} J.G. Amar, F.E. Sullivan, and R.D. Mountain, \textit{Phys. Rev. B}, 1988, \textbf{37}, 196.
		\bibitem{majum_6} S. Majumder and S.K. Das, \textit{Phys. Rev. E}, 2011, \textbf{84}, 021110.
	\bibitem{binder2_6} K. Binder and D. Stauffer, \textit{Phys. Rev. Lett.}, 1974, \textbf{33}, 1006.
	\bibitem{binder3_6} K. Binder, \textit{Phys. Rev. B}, 1977, \textbf{15}, 4425.
	\bibitem{siggia_6} E.D. Siggia, \textit{Phys. Rev. A}, 1979, \textbf{20}, 595.
	\bibitem{furu1_6} H. Furukawa, \textit{Phys. Rev. A}, 1985, \textbf{31}, 1103.
	\bibitem{furu2_6} H. Furukawa, \textit{Phys. Rev. A}, 1987, \textbf{36}, 2288.
	\bibitem{roy_6} S. Roy and S.K. Das, \textit{Phys. Rev. E}, 2012, \textbf{85}, 050602.
	\bibitem{royp_6} S. Roy and S.K. Das, \textit{Soft Matter}, 2013, \textbf{9}, 4178.
	\bibitem{shimi_6} R. Shimizu and H. Tanaka, \textit{Nature Comm.}, 2015, \textbf{6}, 7407.
	\bibitem{mid1_6} J. Midya and S.K. Das, \textit{Phys. Rev. Lett.}, 2017, \textbf{118}, 165701.
	\bibitem{marche_6} M. C. Marchetti, J. F. Joanny, S. Ramaswamy, T. B. Liverpool, J. Prost, 
	M. Rao, and A. Simha, \textit{Rev. Mod. Phys.}, 2013, \textbf{85}, 1143.
	\bibitem{rama1_6} S. Ramaswamy, \textit{Annu. Rev. Cond. Mat. Phys.}, 2010, \textbf{1}, 323.
	\bibitem{cates1_6} M.E. Cates and J.Tailleur, \textit{Annu. Rev. Cond. Mat. Phys.}, 2015, \textbf{6}, 219.
	\bibitem{mishra1_6} S. Mishra and S. Ramaswamy, \textit{Phys. Rev. Lett.}, 2006, \textbf{97}, 090602.
	\bibitem{belmon_6} J.M. Belmonte, G.L. Thomas, L.G. Brunnet, R.M.C. de Almeida, and H. Chat\'{e}, \textit{Phys. Rev. Lett.}, 2008, \textbf{100}, 248702.
	\bibitem{mishra2_6} S. Mishra, A. Baskaran, and M.C. Marchetti, \textit{Phys. Rev. E}, 2010, \textbf{81}, 061916.
	\bibitem{cates2_6} M.E. Cates, D. Marrenduzzo, I. Pagonabarraga, and J. Tailleur, \textit{Proc. Natl. Acad. Sci. U.S.A.}, 2010, \textbf{107}, 11715.
	\bibitem{hagan1_6} G.S. Redner, M.F. Hagan, and A. Baskaran, \textit{Phys. Rev. Lett.}, 2013, \textbf{110}, 055701.
	\bibitem{hagan2_6} G.S. Redner, A. Baskaran, and M.F. Hagan, \textit{Phys. Rev. E}, 2013, \textbf{88}, 012305.
	\bibitem{wys_6} A. Wysocki, R.G. Winkler, and G. Gompper, \textit{Europhys. Lett.}, 2014, \textbf{105}, 48004.
	\bibitem{mehes_6} E. M\'{e}hes, E. Mones, V. N\'{e}meth, and T. Vicsek, \textit{PLOS ONE}, 2012, \textbf{7}, e31711.
	\bibitem{peru_6} F. Peruani and M. B\"{a}r, \textit{New J. Phys.}, 2013, \textbf{15}, 065009.
	\bibitem{mishra3_6} S. Mishra, S. Puri, and S. Ramaswamy, \textit{Phil. Trans. R. Soc. A}, 2014, \textbf{372}, 20130364.
	\bibitem{cremer_6} P. Cremer and H. L\"{o}wen, \textit{Phys. Rev. E}, 2014, \textbf{89}, 022307.
	\bibitem{schwarz_6} J. Schwarz-Linek, C. Valeriani, A. Cacciuto, M. E. Cates, D. Marenduzzo, A.N. Morozov, 
	and W.C.K. Poon, \textit{Proc. Natl. Acad. Sci. U.S.A.}, 2012, \textbf{109}, 4052.
	\bibitem{pala_6} J. Palacci, S. Sacanna, A.P. Steinberg, D.J. Pine, and P.M. Chaikin, \textit{Science}, 2013, \textbf{339}, 936.
	\bibitem{kumar1} N. Kumar, H. Soni, S. Ramaswamy, and A.K. Sood, \textit{Nature Communications}, 2014, \textbf{5}, 4688.
	\bibitem{das1_6} S.K. Das, S.A. Egorov, B. Trefz, P. Virnau, and K. Binder, \textit{Phys. Rev. Lett.}, 2014, \textbf{112}, 198301.
	\bibitem{tref_6} B. Trefz, S.K. Das, S.A. Egorov, P. Virnau, and K. Binder, \textit{J. Chem. Phys.}, 2016, \textbf{144}, 144902.
	\bibitem{vic1_6} T. Vicsek, A. Czir\^{o}k, E. Ben-Jacob, I. Cohen, and O. Schochet, \textit{Phys. Rev. Lett.}, 1995, \textbf{75}, 1226.
	\bibitem{czi_6} A. Czir\^{o}k and T. Vicsek, \textit{Phys. A}, 2000, \textbf{281}, 17.
	\bibitem{bagl_6} G. Baglietto, E.V. Albano, and J. Candia, \textit{Interface Focus}, 2012, \textbf{2}, 708.
	\bibitem{chate_6} H. Chat\'{e}, F. Ginelli, G. Gr\'{e}goire, F. Peruani, and F. Raynand, \textit{Eur. Phys. J. B}, 2008, \textbf{64}, 451.
	\bibitem{das2_6} S.K. Das, \textit{J. Chem. Phys.}, 2017, \textbf{146}, 044902.
		\bibitem{schak_6} S. Chakraborty and S.K. Das, \textit{J. Chem. Phys.}, 2020, \textbf{153}, 044905.
		\bibitem{capri_6} L. Caprini, U.M.B. Marconi, and A. Puglisi, \textit{Phys. Rev. Lett.}, 2020, \textbf{124}, 078001.
	\bibitem{loi_6} D. Loi, S. Mossa, and L.F. Cugliandolo, \textit{Soft Matter}, 2011, \textbf{7}, 10193.
	\bibitem{gold1} I. Goldhirsch and G. Zannetti, \textit{Phys. Rev. Lett.}, 1993, \textbf{70}, 1619.
	\bibitem{yeung1} C. Yeung, \textit{Phys. Rev. Lett.}, 1988, \textbf{61}, 1135.
	\bibitem{mid2_6} J. Midya and S.K. Das, \textit{J. Chem. Phys.}, 2017, \textbf{146}, 024503.
	\bibitem{frenkel_6} D. Frenkel and B. Smit, \textit{Understanding Molecular Simulations: From Algorithm to Applications}, Academic Press, California, 2002.
	\bibitem{allen_6} M.P. Allen and D.J. Tildesley, \textit{Computer Simulations of Liquids}, Clarendon, Oxford, 1987.
	\bibitem{han_6} J.-P. Hansen and I.R. McDonald, \textit{Theory of Simple Liquids}, Academic press, London, 2008.
		\bibitem{loi1_6} D. Loi, S. Mossa, and L.F. Cugliandolo \textit{Phys. Rev. E}, 2008, \textbf{77}, 051111.
	\bibitem{paul2} S. Paul and S. K. Das, \textit{Phys. Rev. E}, 2017, \textbf{96}, 012105.
	\bibitem{paul3} S. Paul and S. K. Das, \textit{Phys. Rev. E}, 2018, \textbf{97}, 032902.
	\bibitem{gold_6} H. Goldstein, C.P. Poole, and J.F. Safko, \textit{Classical Mechanics}, 3rd Ed., Addison-Wesley, 2001.
	\bibitem{vic2_6} T. Vicsek, \textit{Fractal Growth Phenomena}, World Scientific, Singapore, 1992.
	\bibitem{vic3_6} T. Vicsek, M. Shlesinger, and M. Matushita, editors, \textit{Fractals in Natural Sciences}, World Scientific, Singapore, 1994.
	\bibitem{scior1_6} F. Sciortino and P. Tartaglia, \textit{Phys. Rev. Lett.}, 1995, \textbf{74}, 282.
	\bibitem{scior2_6} F. Sciortino, A. Belloni, and P. Tartaglia, \textit{Phys. Rev. E}, 1995, \textbf{52}, 4068.
	\bibitem{huse_6} D.A. Huse, \textit{Phys. Rev. B}, 1996, \textbf{34}, 7845.
	\bibitem{sengers} S. K. Das, J. V. Sengers and M. E. Fisher, \textit{J. Chem. Phys.}, 2007, \textbf{127}, 144506.
	\bibitem{carne_6} G.F. Carnevale, Y. Pomeau, and W.R. Young, \textit{Phys. Rev. Lett.}, 1990, \textbf{64}, 2913.
	\bibitem{trizac1_6} E. Trizac and P.L. Krapivsky, \textit{Phys. Rev. Lett.}, 2003, \textbf{91}, 218302.
	\bibitem{trizac2_6} E. Trizac and J.-P. Hansen, \textit{J. Stat. Phys.}, 1996, \textbf{82}, 1345.
		\bibitem{sten_6} J. Stenhammer, D. Marenduzzo, R.J. Allen, and M.E. Cates, \textit{Soft Matter}, 2014, \textbf{10}, 1489.
		\bibitem{mani_6} E. Mani and H. L\"{o}wen, \textit{Phys. Rev. E}, 2015, \textbf{92}, 032301.
	
\end{thebibliography}
\end{document}